\documentclass{pasa}%

\title[Magnetic equilbrium models]{The impact of superconductivity and the Hall effect in models of magnetized neutron stars}

\author[Sur \& Haskell]{Ankan Sur\thanks{corresponding author:ankansur@camk.edu.pl}\, and Brynmor Haskell
\affil{Nicolaus Copernicus Astronomical Center, Polish Academy of Sciences, Bartycka 18, 00-716 Warsaw, Poland}%

}%

\jid{PASA}
\doi{10.1017/pas.\the\year.xxx}
\jyear{\the\year}

\usepackage{aas_macros}
\usepackage{hyperref} 
\hypersetup{colorlinks,citecolor=blue,linkcolor=blue,urlcolor=blue}

\usepackage{graphicx}
\usepackage{xcolor}
\usepackage{caption}
\usepackage{multirow}
\usepackage{txfonts}
\usepackage[]{hyperref}
\usepackage{subcaption}

\hypersetup{colorlinks,citecolor=blue}
\usepackage{threeparttable}
\usepackage{epstopdf}
\graphicspath{{./Figures/}}
\hypersetup{draft}

\begin{document}

\begin{frontmatter}
\maketitle

\begin{abstract}
{Equilibrium configurations of the internal magnetic field of a pulsar play a key role in modeling astrophysical phenomena, from glitches to gravitational wave emission. In this paper we present a numerical scheme for solving the Grad-Shafranov equation and calculating equilibrium configurations of pulsars, accounting for superconductivity in the core of the neutron star, and for the Hall effect in the crust of the star. Our numerical code uses a finite-difference method in which the source term appearing in the Grad-Shafranov equation, used to model the magnetic equilibrium is nonlinear. We obtain solutions by linearizing the source and applying an under-relaxation scheme at each step of computation to improve the solver's convergence. We
have developed our code in both C++ and Python, and our numerical algorithm can further be adapted to solve any nonlinear PDEs appearing in other areas of computational astrophysics. We produce mixed toroidal-poloidal field configurations, and extend the portion of parameter space that can be investigated with respect to previous studies. We find that in even in the more extreme cases the magnetic energy in the toroidal component does not exceed approximately 5\% of the total. We also find that if the core of the star is superconducting, the toroidal component is entirely confined to the crust of the star, which has important implications for pulsar glitch models which rely on the presence of a strong toroidal field region in the core of the star, where superfluid vortices pin to superconducting fluxtubes.}
\end{abstract}

\begin{keywords}
neutron stars -- pulsars -- magnetic fields -- numerical
\end{keywords}
\end{frontmatter}

\section{Introduction}

The magnetic field in neutron stars (NSs) varies over a very wide range, with strengths ranging from $10^8$ G in old recycled pulsars all the way upto $10^{15}$ G in magnetars. These values of the magnetic field are generally inferred from the observed spindown rate of the star, assuming that it is due to magnetic dipole radiation. This provides information on the exterior field far from the star, however details about the interior field configuration remain unknown. Observationally, the inferred exterior magnetic field of pulsars is found to be relatively stable on short timescales, except for energetic outbursts and flares in magnetars \citep{Rea2011, Coti2018}. Nevertheless, the differences in field strengths between different populations, suggest that on long timescales, comparable to the lifetime of the star, the field may evolve, and that different classes of neutron stars may differ not only due to their age, but also to their magnetic field configuration at birth \citep{Kaspi10}.  Any change in the exterior field is expected to be driven by internal phenomena, such as the flow of currents in the crust and superconductivity in the NS core. The dynamical interplay between these two regions is thus crucial to understand also magnetospheric phenomenology \citep{Akgun2017, KG2014, Kostas2016, Gusakov17}. Understanding the secular evolution of the magnetic field is crucial to connect different evolutionary tracks and NS classes like millisecond pulsars, rotation-powered pulsars and magnetars.  On dynamical timescales, however, it is of interest to understand the physical conditions that allow to obtain stable equilibria in NSs, in order to use such models as backgrounds to model phenomena such as gravitational wave emission due to oscillations or deformations of the crust {\citep{Ushomirsky2000, Payne2006, Osborne2020, Singh20}}.\\

A number of equilibrium models of magnetized stars have been produced in recent years, and to study their stability and evolution, magnetohydrodynamic (MHD) simulations have been performed, all of which have shown to produce quasi-equilibrium mixed poloidal-toroidal geometry starting from the earlier works of 
\citet{2006AA...450.1097B, 2006AA...450.1077B, 2011ApJ...736L...6C} and more recently by \citet{2020MNRAS.495.1360S}. Purely poloidal or purely toroidal magnetic field initial conditions are known to be unstable, and analytical and numerical studies have shown to favor an axisymmteric twisted-torus field \citep{Haskell2008, 2009MNRAS.395.2162L, 2011ApJ...735L..20L, 2012ApJ...760....1C} where the poloidal and the toroidal components stabilizes one other. These models, however, consider a `fluid' star, which makes them relevant only in the first instants of life of the star, when the temperature is too high for the crust to have formed yet. Furthermore in most cases the equation of state is taken to be barotropic, and the stability of barotropic equilibria has been questioned \citep{lander2012,Mitchell2015}. Rotation may provide partial stabilization and in particular the boundary conditions play an important role in determining whether the poloidal or the toroidal field is globally dominant \citep{lander2012}, but stratification provided by charged particles of electrons and protons carrying magnetic flux moving through a neutron fluid in particular, may allow for additional degrees of freedom and allow to stabilize the field \citep{Castillo2017, Castillo2020}. In fact, barotropic axisymmetric equilibrium solutions are unstable under non-axisymmetric perturbations owing to MHD instabilities and stable stratification is likely to be required to prevent complete dissipation of the field \citep{Braithwaite2009, Reisenegger2009, Mitchell2015}. Nevertheless, while non-barotropicity may be crucial to understand the stability of the models, the equilibria themselves will not differ significantly from those of bartoropic stars in mature pulsars \citep{Castillo2020}, making barotropic equilibria an important tool to use in calculations that require magnetized background models of NSs.\\

{It is well known that rotation (or more in general nontrivial fluid-velocity fields in the stellar interior) may have an impact on the evolution of the magnetic field. Firstly, instabilities due to perturbations developed within the NS are stabilized by rotation. Secondly, superfluids in deferentially-rotating NS cores experience torque oscillations \citep{Peralta:2005xw,Melatos2007}, which are likely to explain glitches observed in standard pulsars. Thirdly, internal velocity fields and multifluid components could give rise to additional modes of oscillations and alter the properties of modes of non-rotating stars \citep{Akg2008}. Fourthly, on studying magneto-thermal evolution with macroscopic flux tube drift velocity, it had been shown that magnetic field may be weakly buried in the outermost layers of the core and not completely expelled, as previously thought, although this is sensitive to the initial conditions \citep{Elfritz2016}. And lastly, the presence of bulk motion in the crust was explored in \cite{Kojima2021} who showed that the magnetic energy is converted into mechanical work and parts of it are dissipated through bursts or flares. A realistic model of NS should consider a solid crust and a fluid core, which are in rotation (and possibly in differential rotation due to hydromagnetic torques). However, in this work, we neglect the effects of rotation as we are mainly interested in the magnetic field configuration in mature pulsars, which are slowly rotating, and in understanding the impact of suprconductivity in the core and of the Hall effect in the crust. It is, nevertheless, important to keep in mind that rotation may play an important role in the evolution of younger, strongly magnetized, NSs, and should be considered to obtain a full picture of the evolution of the field during the lifetime of a NS. }\\

The crust of a NS consists of $\approx 1\%$ of the total mass but plays an important role for the dynamics and emission properties of the star. The composition of these outer layers depends on the equation of state and the density varies from $10^{6} \, \rm gm \,cm^{-3}$ in the outer crust to $\sim 10^{14} \,\rm  gm \,cm^{-3}$ at which point there is a transition to a fluid outer core of neutrons, protons, electrons and muons, and at higher densities still, in the inner core, one may have an inner core of exotic particles like hyperons, superconducting quark matter and Boson condensates. When the temperature drops below $T\approx 10^9$ K, soon after birth, the crusts begins to solidify, and forms conducting crystal lattice with free electrons soaked in superfluid neutrons where the Lorentz force can be balanced by elastic forces. During the lifetime of a NS, the evolution of the field in the crust is mainly affected by two processes: (a) the Hall effect, and (b) Ohmic dissipation, owing to the currents carried by electrons in the crust \citep{Goldreich1992, Cumming2004, Pons2007, 2002MNRAS.337..216H}. It is known that Hall effect leads to turbulent cascades but whether it leads to complete dissipation of the field or relaxes to a stable state is an important question as stationary closed configuration is neutrally stable \cite{Lyutikov}. Over the Hall timescale, \citep{KSAC2014} have shown that indeed the field evolves to a state known as the ``Hall attractor'' having a dipolar poloidal field and a weak quadrupolar toroidal component. Depending on the steepness of the electron density, this field may dissipate rapidly \citep{2013MNRAS.434.2480G}. As the field relaxes from an MHD (fluid) to a Hall equilibrium, it may drive the expulsion of toroidal loops powering flares from the NS crust \citep{Thompson1995}.\\

When the temperature drops below $\approx 10^9$ K in the core, the protons will be superconducting and the neutrons superfluid \citep{HaskellSedrakian}, which will have a significant effect on the evolution of the field in the standard pulsar population \citep{Ofengeim2018, Gusakov2017, Gusakov2020}. Superconductivity, in particular, affect the magnetic field, as if it is of type II, as theoretical models suggest, the field will be confined to flux tubes, which can also interact with superfluid neutron vortices (see \citealt{HaskellSedrakian} for a review). In fact, the possibility that neutron vortices may pin in strong toroidal field regions in the superconducting core has been proposed as an explanation for the observed high values for the activity parameter in glitching pulsars such as the Vela \citep{Erbil14, Erbil17, Erbil20}.
 In the core, \cite{Goldreich1992} was the first to propose that ambipolar diffusion becomes important where the charged particles like electrons and protons move relative to the neutrons. \cite{Kostas2011} showed that this ambipolar diffusion in superconducting/superfluid NSs has negligible effect on the magnetic field evolution. However this can change if the core-temperature is of the order $10^8-10^9$ K and the diffusion time scale is comparable to age of the star \cite{Passamonti2017}.\\

A realistic model for the magnetic field structure of a standard pulsar cannot be that of a magnetized fluid star, and thus MHD equilibrium, but should include a superconducting core and a crust. In this paper, we therefore construct equilibrium models for magnetized NS, including type-II superconductivity in the core and the Hall effect in the crust, and compare our models to pure MHD and Hall equilibria.\\

The equilibrium of the magnetic field is studied by solving the so-called Grad-Shafranov (GS) equation \citep{Shaf1966}, whose formalism we discuss in the next section. This GS equation appears widely in plasma physics and analytical solutions are often hard to obtain. Nevertheless, when the source term has a simple form, we can use Green's functions to solve the GS equation. However, except for a small number of simple forms for the source function, numerical methods such as finite differences \citep{Johnson79}, spectral methods \citep{Ling85}, spectral elements \citep{Howell2014} and linear finite elements \citep{Gruber1987} should be used. In applications to NSs, numerical solvers such as the HSCF method \citep{2009MNRAS.395.2162L}, Gauss-Seidel method \citep{2013MNRAS.434.2480G} or the generalized Newton's method  \citep{2015ApJ...802..121A} have been used.\\

We propose a numerical technique based on finite-difference iterative scheme for solving the GS equation. We focus in the astrophysical relevance of the GS equation, in particular, to obtain magnetic equilibrium configurations in neutron stars. {Our method is fast, written both in C++ and python, and easier to implement numerically. We have generated models in regimes where numerical instabilities were faced by previous works. In order to do this, we have demonstrated how nonlinear source terms can be treated numerically for the first time}. Our numerical algorithm in general can be applied to any such nonlinear PDEs of the similar form, like the \textit{Poisson} equation, appearing ubiquitously in physics.\\

This article is arranged as following: in section 2 we derive the GS equation for Hall and MHD equilibrium, in section 3, we describe the numerical algorithm to solve the discretized GS equation, in section 4 we show our results for the normal matter star and the superconducting core, while conclusions and discussions are finally presented in section 5. 

\begin{figure*}
	\centering
	\includegraphics[scale=0.6]{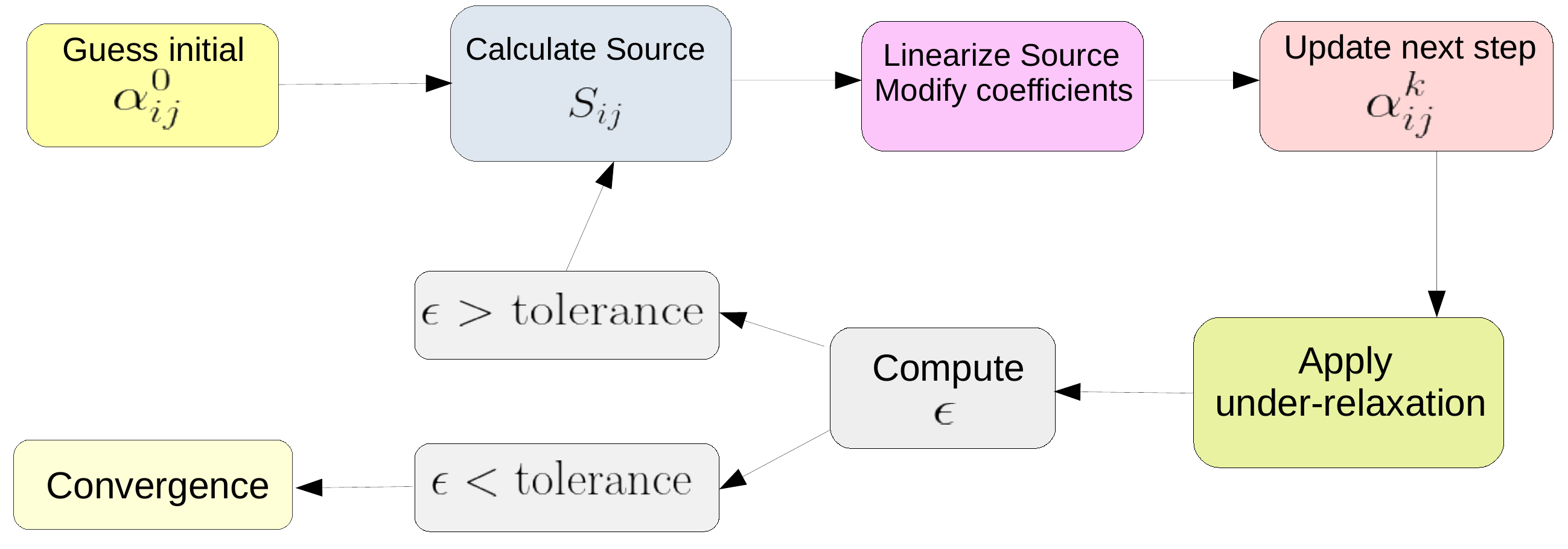}
	\caption{Flowchart of our numerical algorithm.}
	\label{flwchart}
\end{figure*}

\section{Mathematical formalism}
{In general, the magnetic field $(\vec{B})$ in spherical coordinates is expressed in terms of two scalar functions, $\alpha(r,\theta)$ representing the poloidal component, and $\beta(r,\theta)$ representing the toroidal component, as 
\begin{eqnarray}
\vec{B} = \vec{\nabla}{\alpha}\times\vec{\nabla}\phi + \beta\vec{\nabla}\phi
\label{eq_ini}
\end{eqnarray}
where $\vec{\nabla}\phi = \hat{\phi}/r\sin\theta$. From Faraday's law, we have 
\begin{eqnarray}
\frac{\partial \vec{B}}{\partial t} = -c\vec{\nabla} \times \vec{E}
\label{Faraday}
\end{eqnarray}
where the electric field is $\vec{E} = -\frac{1}{c}\vec{v}\times\vec{B} + \frac{\vec{j}}{\sigma}$, $\vec{v}$ is the velocity of electrons which is related to the current density as $\vec{v} = -\frac{\vec{j}}{e\,n}$, $n$ is the electron density and $\sigma$ is the electrical conductivity. Moreover, Ampere's law states that the current density is related to the magnetic field as $\vec{j}=\frac{c}{4\pi}\vec{\nabla}\times \vec{B}$, and substituting these in equation (\ref{Faraday}) yields the induction equation,
\begin{eqnarray}
\frac{\partial \vec{B}}{\partial t} = -\frac{c}{4\pi e}\vec{\nabla}\times\bigg(\frac{\vec{\nabla}\times \vec{B}}{n} \times \vec{B}\bigg) - \frac{c^2}{4\pi}\vec{\nabla} \times \bigg(\frac{\vec{\nabla}\times \vec{B}}{\sigma}\bigg)
\label{ind}
\end{eqnarray}
{The first term on the right hand side of the above equation is referred to as the Hall term while the second is the Ohmic dissipation term. The ratio of timescales on which these two terms operate is given by \citep{Goldreich1992}
\begin{eqnarray}
\frac{\tau_{\rm Ohm}}{\tau_{\rm Hall}} = 4\times10^{4}\frac{B_{14}}{T_8^2}\bigg(\frac{\rho}{\rho_{\rm nuc}}\bigg)^2
\end{eqnarray}
where $B = B/10^{14}\,\rm G$ and $T_8 = T/10^8 \rm K$. Thus for a suitable choice of the parameters density, magnetic field, and temperature, the Hall effect is faster than the Ohmic term and we can obtain a family of Hall equilibrium solutions. In particular we expect this to be true in the cores of standard pulsars, with $B\approx 10^{12}$ G and internal temperatures of the order of $T\approx 10^7$ K.}

The evolution of the magnetic field purely due to Hall effect is given by 
\begin{eqnarray}
\frac{\partial \vec{B}}{\partial t} = -\frac{c}{4\pi e}\vec{\nabla}\times\bigg(\frac{\vec{\nabla}\times \vec{B}}{n} \times \vec{B}\bigg)
\label{Halleq}
\end{eqnarray}
To obtain steady-state models, axisymmetric Hall equilibria solutions are calculated by setting equation \ref{Halleq} to zero. Integrating this equation gives
\begin{eqnarray}
\frac{1}{n}(\vec{\nabla} \times \vec{B})\times \vec{B} = \vec{\nabla} \chi_{\rm Hall}
\label{eq_hall}
\end{eqnarray}
where $\chi_{\rm Hall}$ is an arbitrary function of the coordinates $r$ and $\theta$, which can be physically interpreted as the magnetic potential since its gradient gives the magnetic force. Substituting equation \ref{eq_ini} gives the toroidal component as
\begin{eqnarray}
\vec{\nabla}\alpha \times \vec{\nabla}\vec{\beta} = 0
\end{eqnarray} 
which shows $\beta = \beta(\alpha)$. Moreover, $\vec{\nabla}\alpha \parallel \vec{\nabla}\chi_{\rm Hall}$ implies $\chi_{\rm Hall} = \chi_{\rm Hall}(\alpha)$. This gives rise to the Grad-Shafranov (GS) equation for a two-dimensional plasma, which is a second-order nonlinear partial differential equation (PDE) given by:
\begin{equation}
\Delta^{\star} \alpha = \frac{\partial^2 \alpha}{\partial r^2} + \frac{(1-\mu^2)}{r^2}\frac{\partial^2 \alpha}{\partial \mu^2} = -\chi^{\prime}(\alpha)n(r) r^2(1-\mu^2)  - \beta^{\prime}\beta = -\mathcal{S}
\label{eq1}
\end{equation}
where $\Delta^{\star}$ is the GS operator, $\mu=\cos(\theta)$ and $\mathcal{S}$ is the source term.

The GS equation, however, does not only apply to Hall equilibria. In a barotropic NS, i.e. where the pressure is a function of mass density ($\rho$) alone, as $P=P(\rho)$, a very similar form of equation \ref{eq_hall} is also obtained for MHD equilibria, for which one has:
\begin{equation}
\frac{1}{4\pi \rho}(\vec{\nabla} \times \vec{B})\times \vec{B}=\frac{\nabla p}{\rho}+\nabla{\phi}
\label{MHDfull}
\end{equation}
with $\phi$ the gravitational potential. For a barotropic equation of state, equation (\ref{MHDfull}) is clearly of the same form as the GS equation, and can thus be written in the same form as (\ref{eq_hall}):
\begin{eqnarray}
\frac{1}{\rho}(\vec{\nabla} \times \vec{B})\times \vec{B} = \vec{\nabla} \chi_{\rm MHD}
\end{eqnarray}	
where, however, the specific terms have different interpretations with respect to Hall equilibria \citep{2013MNRAS.434.2480G}. Specifically in MHD the mass density plays a similar role as the electron density while $\chi_{\rm MHD}$ as $\chi_{\rm Hall}$. The poloidal field evolution takes the same form as equation \ref{eq1}, and thus it is necessary to obtain a numerical solution in either case of Hall or MHD equilibrium. In this study, we neglect any relativistic terms and assume that the conductivity is high enough that we can neglect the contributions of the Ohmic dissipation term, which is a good approximation in NS interiors. {However, it should be noted that the Ohmic dissipation term is likely to be present in the crust and the Hall drift enhances it by  forming small-scale eddies through which the field dissipates magnetic energy \citep{Goldreich1992}. Simulations have shown that the Hall drift term quickly saturates and the evolution of the field occurs on a slower Ohmic timescale \citep{Pons2010,Kojima2012,Vigano2012}. Even if the field evolves rapidly during the initial stages, it approaches one among the family of steady-state Hall equilibrium solutions. Given that diffusivity only depends on radius, the Ohmic term will not affect the angular structure of the magnetic field. \cite{2013MNRAS.434.2480G} showed that the electron fluid in the crust slows down rapidly compared to the Ohmic dissipation rate for a field connecting an external dipole. Further, the Hall term enhances the dissipation rate of higher order Ohmic modes as compared to pure Ohmic decay. To fully investigate the evolution of the field, one must solve the induction equation given in (\ref{ind}) as carried out by \cite{marchant2014} who showed that starting from  either purely poloidal equilibrium or an unstable equilibrium initial condition, the Ohmic dissipation evolved the field towards an attractor state through adjacent stable configurations superimposed by damped oscillations. Considering the effects of Ohmic term in our calculations is beyond the scope of this work, and we assume that as the Ohmic decay occurs on a much larger timescale as compared to the Hall timescale, our equilibria are an adequate approximation to the field configurations in a middle aged pulsar.} \\

\section{Numerical Method}

In this section, we discuss our finite-difference iterative scheme for solving the GS equation in spherical coordinates. We consider a two dimensional grid on $r-\mu$ plane. There are $N_r$ points in the $r$ direction running from $r=r_{min}$ at $i=0$ to $r=r_{max}$ at $i=N_r-1$. The radial values have been normalized by the radius of the star (R) so that $r=1$ corresponds to the stellar surface. Similarly, in the $\mu$ direction, we have $N_{\mu}$ points running from $\mu=-1$ at $j=0$ and $\mu=+1$ at $j=N_{\mu}-1$. 
The source term $(S)$ is given by
\begin{align}
\mathcal{S} = 
\begin{cases}
\chi^{\prime}(\alpha)n(r) r^2(1-\mu^2)  + \beta^{\prime}\beta       & \quad \text{if } r  < 1\\
0  &\quad \text{if } r\geq1
\end{cases}
\end{align}
where $\beta$ is the toroidal component and the prime denotes derivative with respect to $\alpha$. {The functional form of $\beta$ does not allow toroidal currents outside the star and thus makes the toroidal field to be located within the stellar interior}. The electron density is assumed to be isotropic within the star, implying $n=n(r)$, and is zero outside due to vacuum. This makes the source term also zero outside the stellar surface. Moreover, this electron density appearing in Hall equilibria states are related to MHD equilibria by $n = \rho Y_e$, where $Y_e$ is {the electron number per unit mass which varies from $10^{22}-10^{28} \, \rm gm^{-1}$ across the crust.} \\

We apply second-order finite difference scheme for equation \ref{eq1} on a two-dimensional grid of $(r-\mu)$ 
\begin{equation}
\frac{\alpha_{i+1,j} + \alpha_{i-1,j} - 2\alpha_{ij}}{dr^2} + \frac{(1-\mu_j^2)}{r_i^2}\frac{\alpha_{i,j+1} + \alpha_{i,j-1} - 2\alpha_{ij}}{d\mu^2} = - \mathcal{S}_{ij} = \mathcal{Q}_{ij}
\end{equation}
where $dr = (r_{max}-r_{min})/(N_r-1)$, $d\mu=2/(N_{\mu}-1)$ and $\mathcal{Q}$ is the negative value of the source function $\mathcal{S}$. On rearranging the above terms, we can get an expression for $\alpha_{ij}$ at the $(k)$-th step in terms of all its neighboring points,
\begin{equation}
\alpha_{ij}^{k} = \frac{(\alpha_{i+1,j}^{k} + \alpha_{i-1,j}^{k})/dr^2}{\omega_{ij}}\\ + \frac{(1-\mu_j^2)}{r_i^2d\mu^2}\frac{\alpha_{i,j+1}^{k} + \alpha_{i,j-1}^{k}}{\omega_{ij}} +  \frac{\mathcal{S}_{ij}}{\omega_{ij}}
\label{iterates}
\end{equation}
where $\omega_{ij} = 2/dr^2 + 2(1-u_j^2)/r_i^2/d\mu^2$. We use updated values of $\alpha_{ij}$ whenever they are available. The boundary conditions were set to $\alpha(r,\mu=-1) = 0$, 
$\alpha(r,\mu=1)=0$, $\alpha(r=r_{min},\mu)=0$, and $\alpha(r=r_{max},\mu)=0$. Axisymmetry equilibrium requires the azimuthal component of the magnetic field to vanish, which allows us to consider a toroidal component of the form
\begin{equation}
\beta = s[\alpha-\alpha(r=1,\mu=0)]^{p}\Theta(\alpha-\alpha(r=1,\mu=0))
\end{equation}
where $s$ and $p$ are free parameters \citep{2009MNRAS.395.2162L,
  2013MNRAS.434.2480G, 2015ApJ...802..121A, 2012MNRAS.422..434F} and
$\Theta$ is the heaviside function. This form ensures there are no
toroidal currents outside the star. The value of $\alpha(1,0)$ is self
consistently calculated at each iteration. This form of $\beta$ also
makes the source term nonlinear. Solving a PDE with non-linear source
terms is a challenging task and we follow the procedure  of \citet{SM2015} to linearize the process. To do so, we expand the source term, namely $\mathcal{Q}$, in Taylor's series
\begin{equation}
\mathcal{Q}_{ij}^{k} = \mathcal{Q}_{ij}^{k-1} + \frac{d\mathcal{Q}}{d\alpha}\biggr \rvert^{k-1} (\alpha_{ij}^{k} - \alpha_{ij}^{k-1})  + .... = \mathcal{Q}_c + \mathcal{Q}_p\alpha_{ij}^{k}
\end{equation}
and neglect the contributions from higher order terms in $\alpha$. We define
\begin{align}
&\mathcal{Q}_c = \mathcal{Q}^{k-1}_{ij} - \frac{d\mathcal{Q}}{d\alpha}\bigr \rvert^{k-1}\alpha_{ij}^{k-1} \\
&\mathcal{Q}_p = \frac{d\mathcal{Q}}{d\alpha}\bigr \rvert^{k-1}
\end{align}
and bring $\mathcal{Q}_p\alpha_{ij}^{k}$ on the left hand side of equation \ref{iterates}, modifying the coefficient $\omega_{ij}$ and the source term such that
\begin{align}
&\omega_{ij} = \omega_{ij} - \text{min}(0, \mathcal{Q}_p) \\
&\mathcal{Q}_{ij} = \mathcal{Q}_{c} + \text{max}(0, \mathcal{Q}_p)\alpha_{ij}^{k-1}
\end{align}
This step does not always guarantee convergence. To improve the performance of our solver, we use an under-relaxation scheme such that
\begin{equation}
\alpha_{ij}^{k} = \xi\alpha_{ij}^{k} + (1-\xi)\alpha_{ij}^{k-1}
\end{equation}
where the parameter $\xi$ lies between 0 and 1. The exact value of $\xi$ depends on the problem and generally a smaller value like 0.1 would make the convergence slower but more accurate. We solve equation \ref{iterates} until a tolerance limit is reached,
\begin{equation}
\frac{\alpha_{ij}^{k}-\alpha_{ij}^{k-1}}{\alpha_{ij}^{k-1}} \leq \varepsilon = 10^{-8}
\end{equation}
where the error at each step is computed as
\begin{equation}
\varepsilon = \sqrt{\frac{\sum \limits_{i,j}(\alpha_{ij}^{k}-\alpha_{ij}^{k-1})^2}{\sum \limits_{i,j} (\alpha_{ij}^{k-1})^2}}
\end{equation}
We have developed Python code\footnote{Our code is freely available for download at \href{https://github.com/ankansur/GSsolver}{https://github.com/ankansur/GSsolver}}, which is a widely used programming language known for its easily available numerical and scientific modules for computing. Instead of looping in the $r-\mu$ plane and calculating each $\alpha_{ij}$ term one at a time, we use \texttt{numpy} \citep{2020arXiv200610256H} vectorization allowing for operations on the entire array at once. This speeds up the code by a factor of 100 for a grid of $50\times50$ and the improvement is more significant for a larger grid size, say $100 \times 100$ or $200 \times 200$. Additionally, we have the same version in C++ which is faster than python and can be used for larger grid-dimensions required to resolve strong toroidal fields.

\section{Results}

Given an EOS for a NS, one can solve the {Tolman}-Oppenheimer-Volkov (TOV) equations to obtain the mass ($M$), radius ($R$) and density profile ($\rho$) of the NS. This information is used as the background model on which we solve the GS equation to obtain the equilibrium magnetic field structure. In this paper, as we are solving Newtonian equations of motion, we use a set of EOSs that are analytically tractable and allow to mimic the the structure of fully relativistic solutions. In particular we produce models with three EOS. First of all we use two particular exact solutions of the Einstein field equations which are of interest for a NS: the first known as the `Schwarzschild' solution gives $\rho = \rho_c = \rm const$ and the other obtained by \cite{Tolman39} gives $\rho(r) = \rho_c(1-r^2/R^2)$, which is close to the density profile for the polytropic EOS $P(\rho) \sim \rho^2$, and has been used previously in several studied of magnetized neutron stars \citep{Mastrano2012,Mastrano2013}. We verify this explicitly by also producing a third set of equilibrium models for an $n=1$ polytropic EOS which is obtained by solving the Lane-Emden equation giving $\rho=\rho_c \frac{sin(\pi r/R)}{\pi r/R}$.This allows us to maintain a physically plausible density profile, and an analytically tractable model, where we can arbitrarily choose $M$ and $R$ to match the compactness predicted by microscopic EOSs, without the technical difficulties associated with the use of a tabulated EOS in our scheme. Unless otherwise stated, we consider as our standard model a NS made of two different regions, a crust of thickness 1 km and a core of thickness 9 km with the crust-core interface having a density $\rho_{cc} \sim 1.9 \times 10^{14}$ gm cm$^{-3}$ and $\rho_c \approx 10^{15}$ gm cm$^{-3}$. All the radial values $r$ appearing hereafter have been normalized by the radius of the star $R$.} 

\begin{figure}
	\centering
	\includegraphics[scale=0.3]{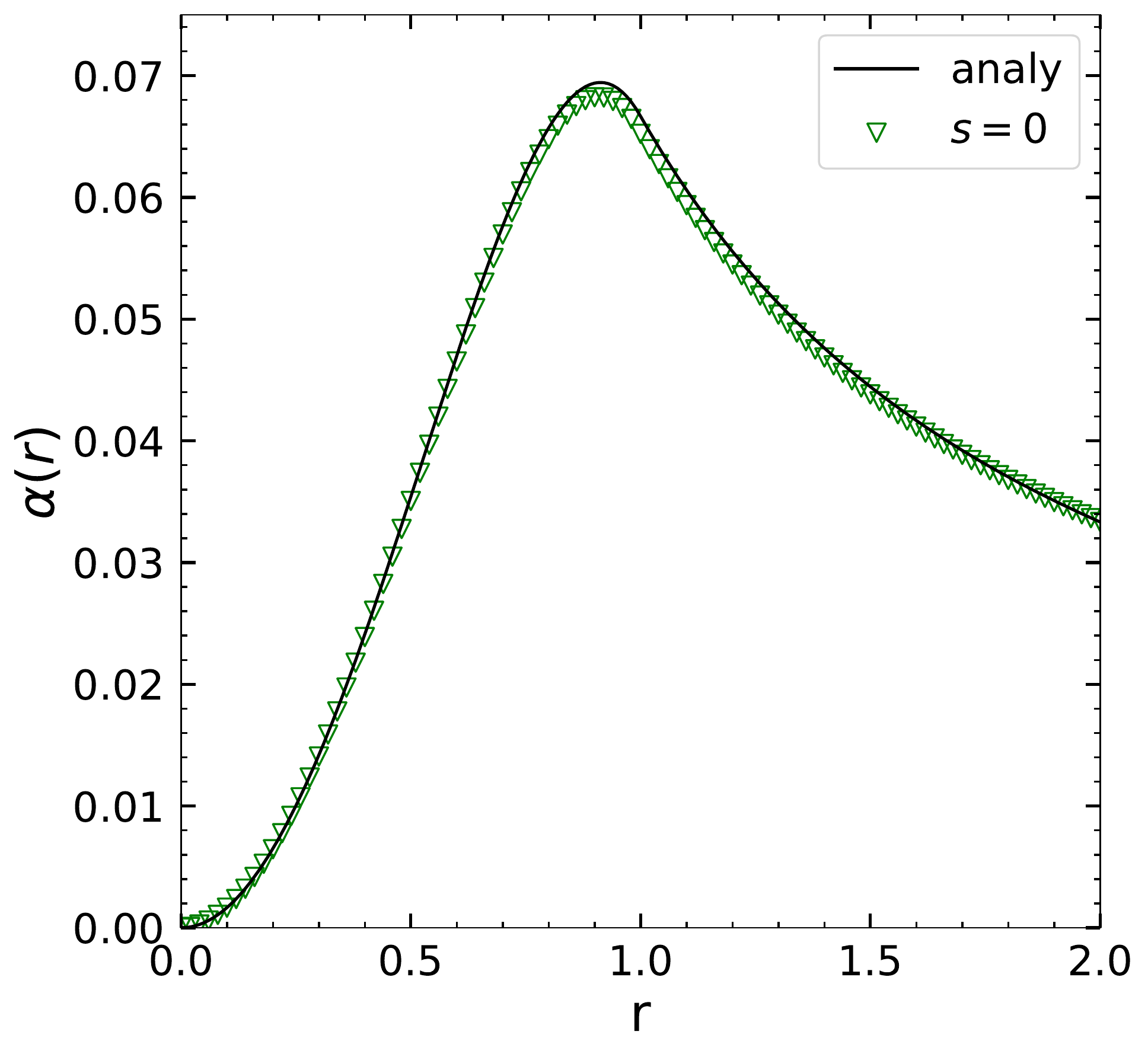}
	\caption{ Variation of $\alpha$ (contours of which give the poloidal field lines) at the equator across the radial direction for the constant electron density profile. The black solid line shows the analytical solution.}
	\label{analy_sol}
\end{figure}

\begin{figure*}
	\centering
	\begin{subfigure}{.5\textwidth}
		\centering
		\includegraphics[scale=0.27]{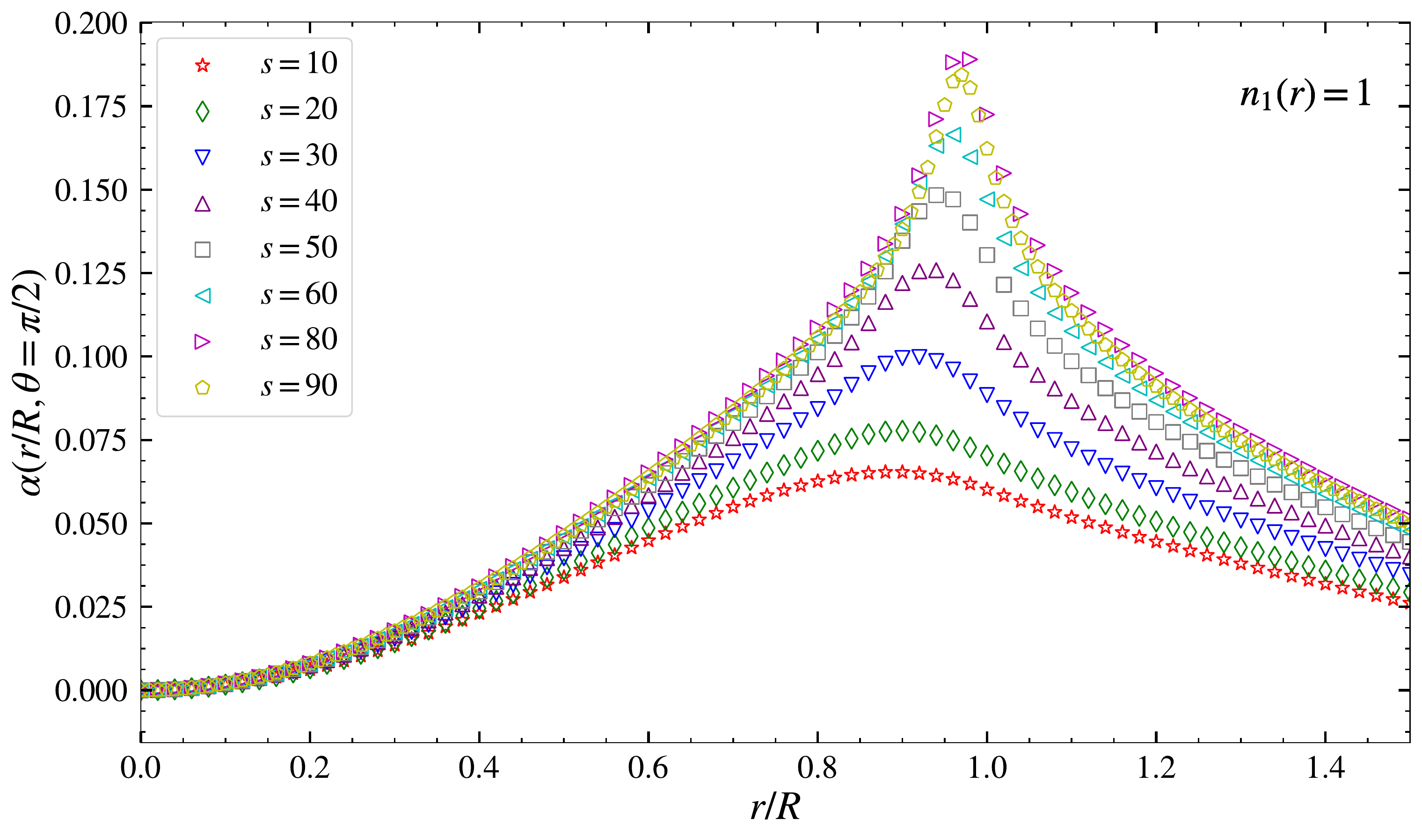}
		\caption{}
		\label{sol1a}
	\end{subfigure}%
	\begin{subfigure}{.5\textwidth}
		\centering
		\includegraphics[scale=0.27]{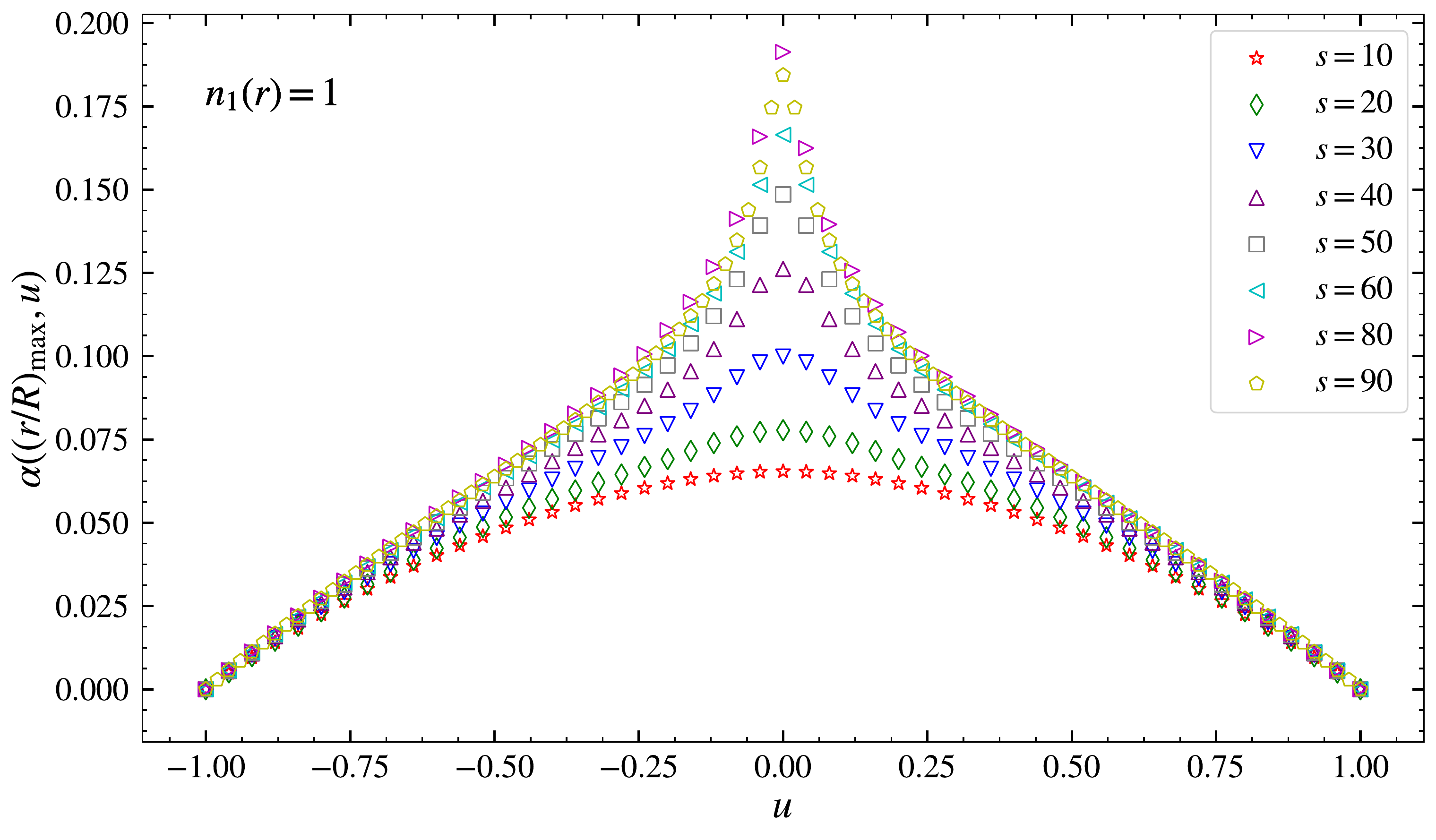}
		\caption{}
		\label{sol1b}
	\end{subfigure}%
	
	\begin{subfigure}{.5\textwidth}
		\centering
		\includegraphics[scale=0.27]{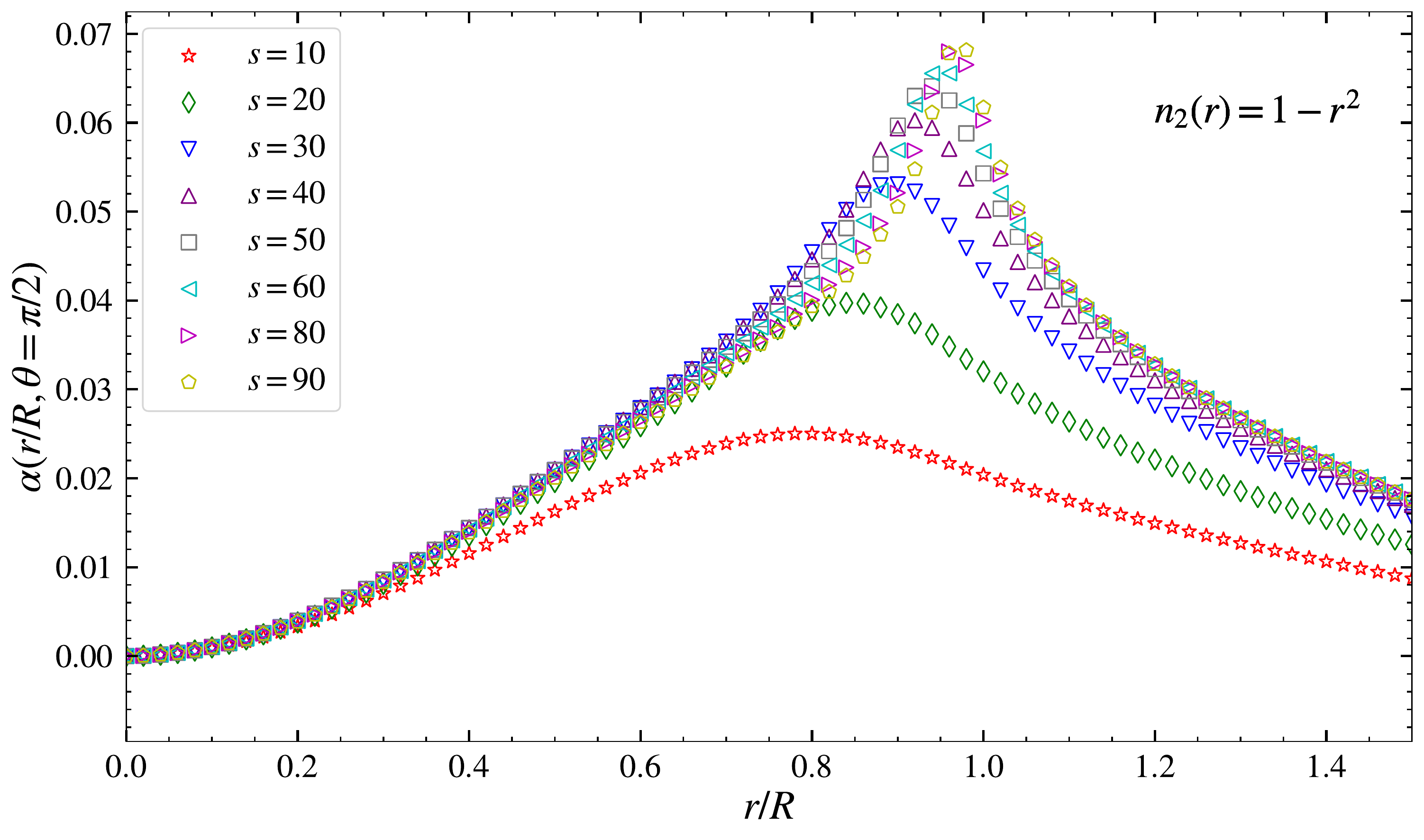}
		\caption{}
		\label{sol2a}
	\end{subfigure}%
	\begin{subfigure}{.5\textwidth}
		\centering
		\includegraphics[scale=0.27]{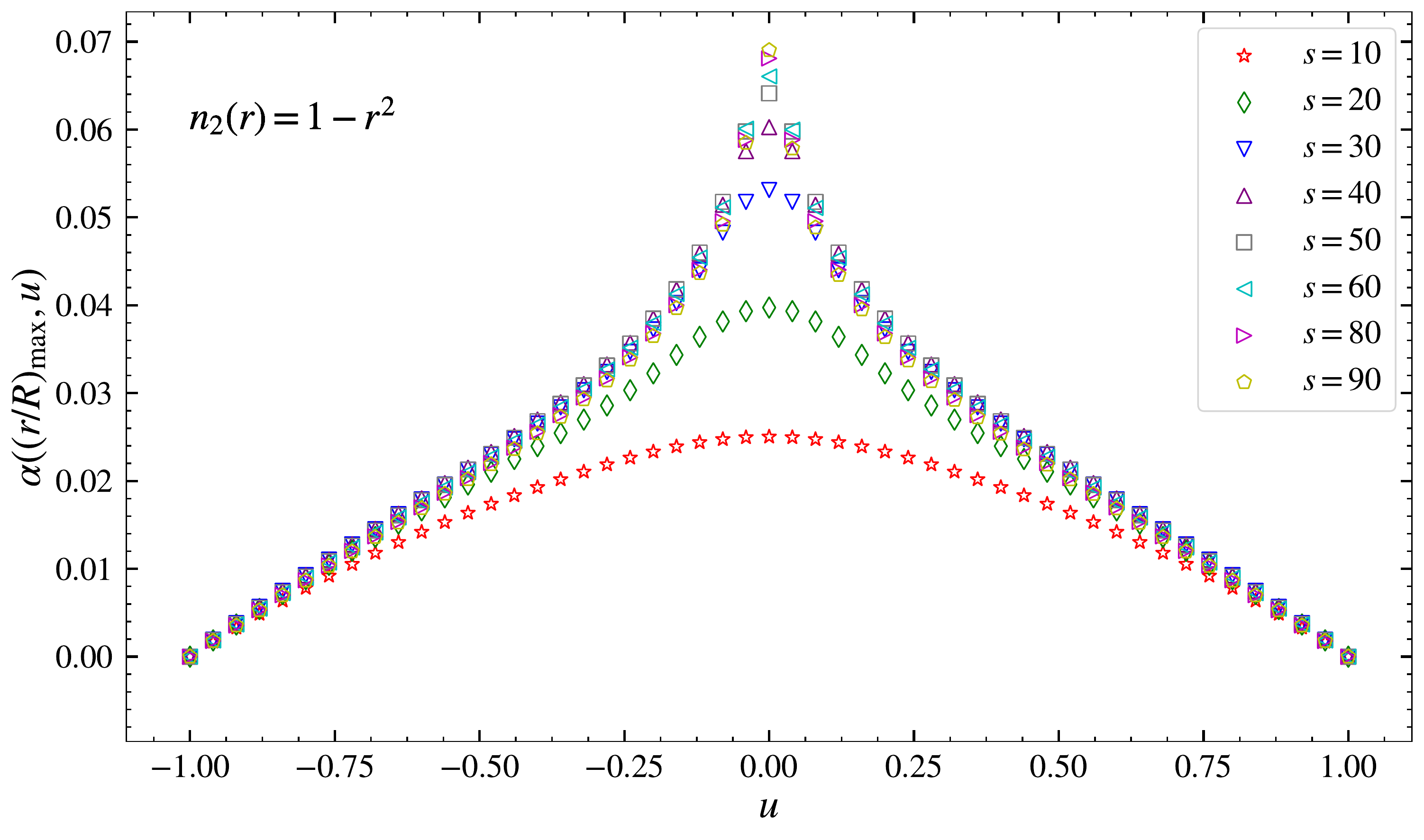}
		\caption{}
		\label{sol2b}
	\end{subfigure} %
	\begin{subfigure}{.5\textwidth}
		\centering
		\includegraphics[scale=0.27]{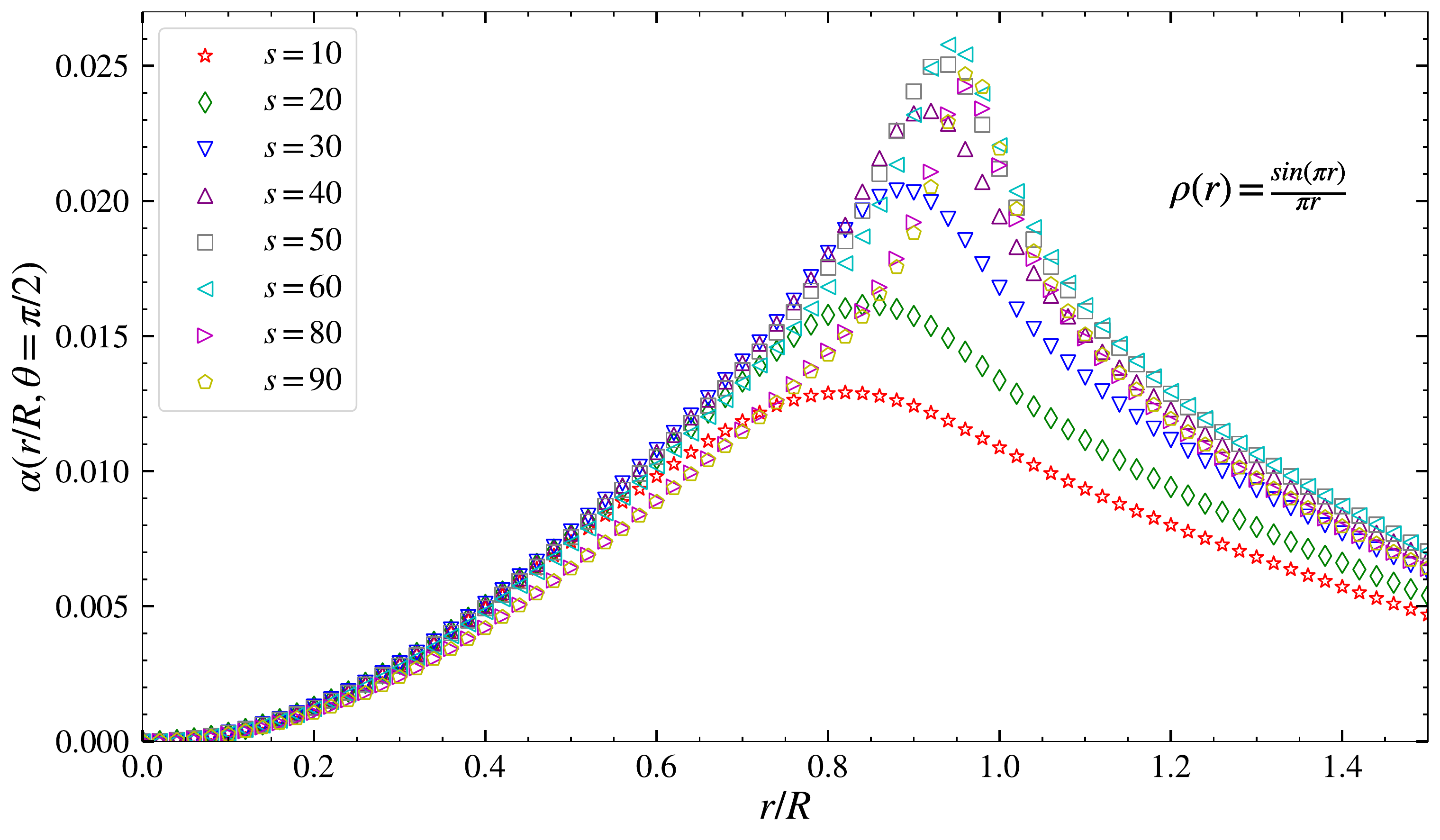}
		\caption{}
		\label{sol3a}
	\end{subfigure}%
	\begin{subfigure}{.5\textwidth}
		\centering
		\includegraphics[scale=0.27]{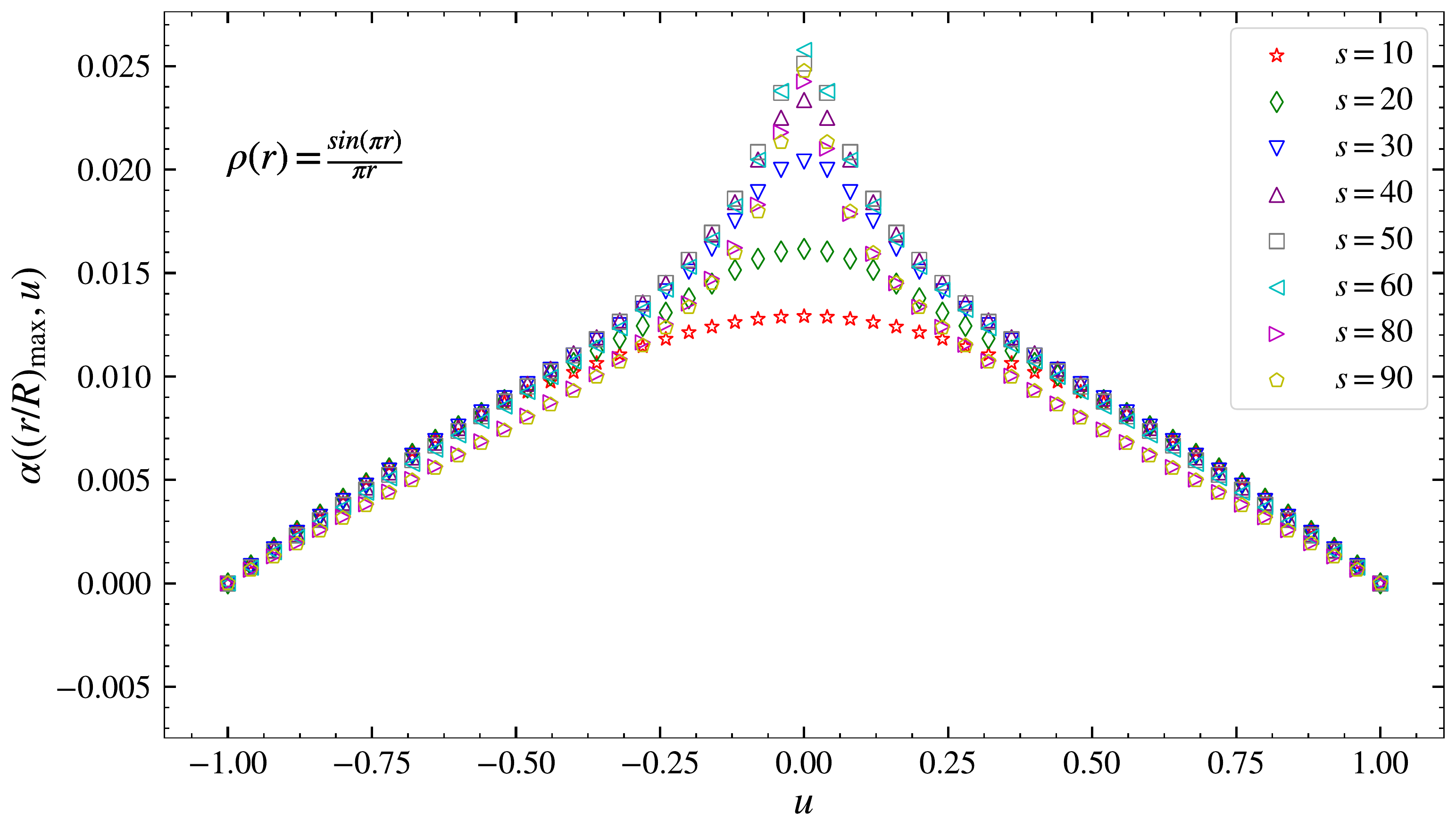}
		\caption{}
		\label{sol3b}
	\end{subfigure} %
	\caption{Variation of $\alpha$ (whose contours give the poloidal field lines) at the equator across the radial direction for 8 values of the parameter
		$s$ shown for three different density profiles in (a),(c) and (e). Variation of the
		maximum value of $\alpha$ across the angular direction with $s$. The density profiles are given as text in each figure. }
	\label{variation_alpha}
\end{figure*}

\begin{figure*}
	\centering
	\includegraphics[scale=0.42]{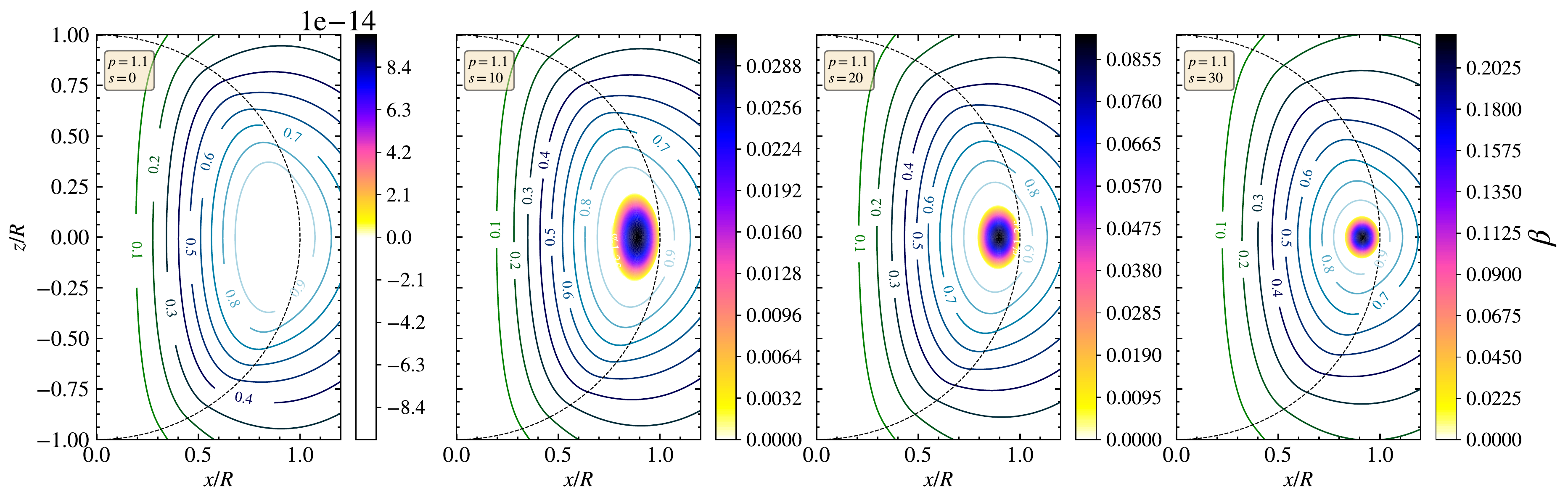}
	\includegraphics[scale=0.42]{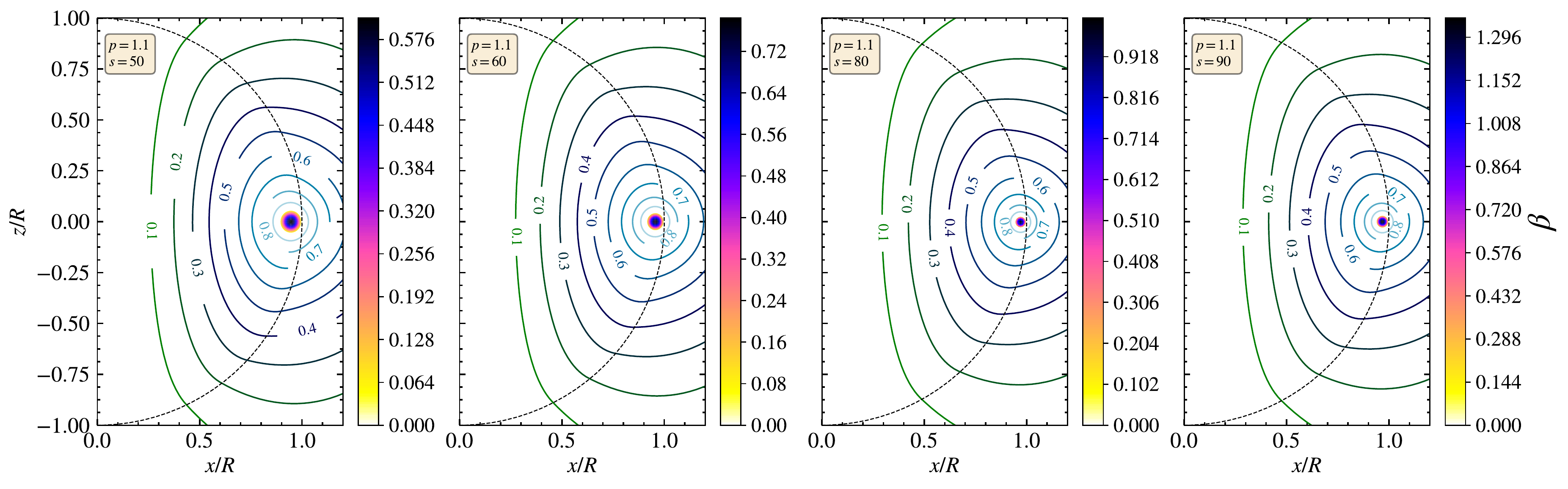}
	\caption[short]{Contours of poloidal field for different values of $s$ (given in the text-box in each figure) using the constant electron density profile $n_1=1$. The colorbar shows the strength of $\beta$. The black dotted line represents the location of stellar radius.  We have shown contours of $\alpha$ having values $(0.1 \alpha_s,0.2\alpha_s,0.3\alpha_s,0.4\alpha_s,0.5\alpha_s,0.6\alpha_s,0.7\alpha_s,0.8\alpha_s,0.9\alpha_s,1\alpha_s)$.}
	\label{contn1}
\end{figure*}

\begin{figure*}
	\centering
	\includegraphics[scale=0.42]{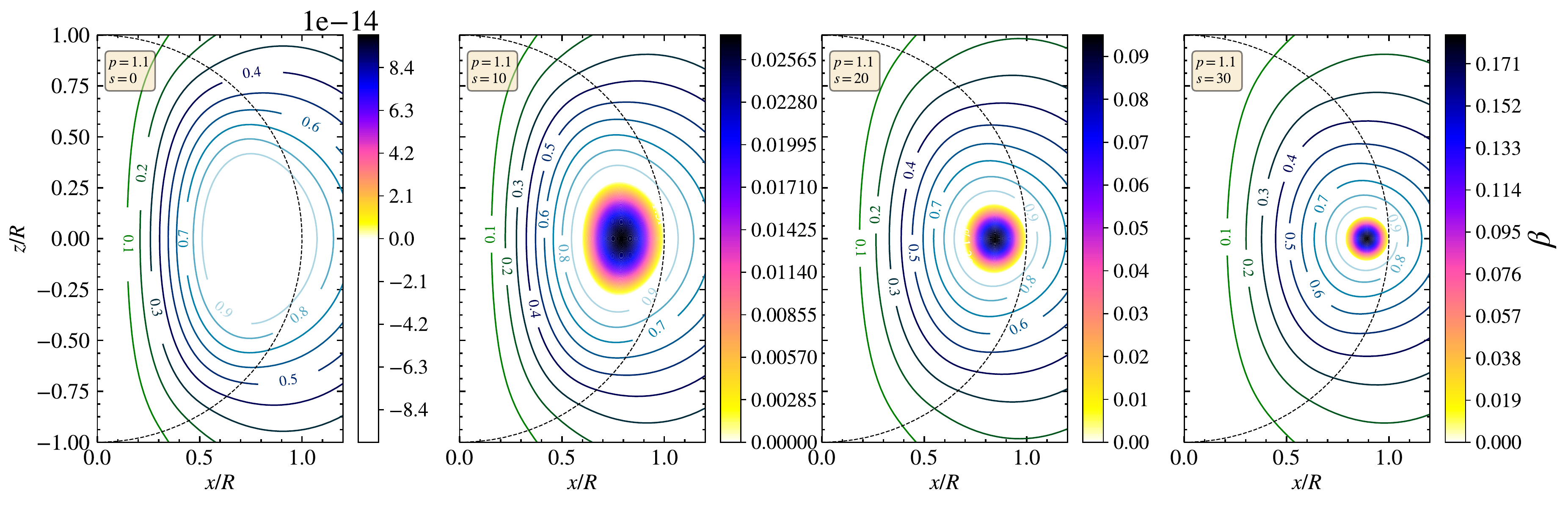}
	\includegraphics[scale=0.42]{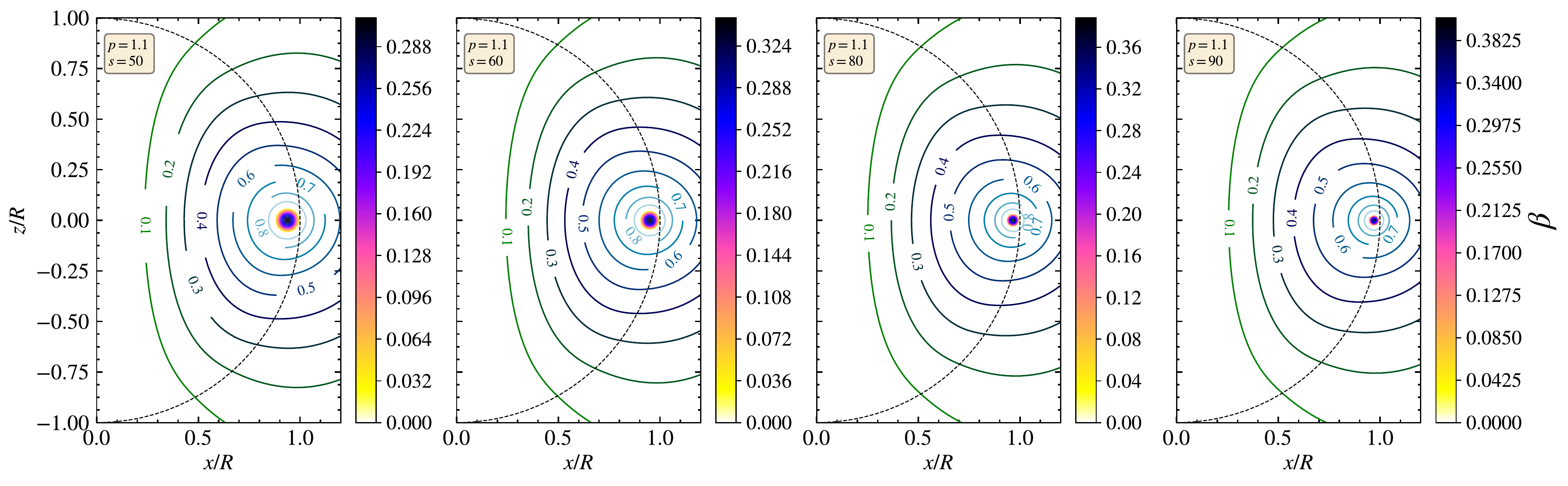}
	\caption[short]{Contours of poloidal field lines (with similar strengths as above), for the electron density profile $n_2=(1-r^2)$. The colorbar, again, shows the strength of $\beta$ and the red dotted line represent $r=R$. }
	\label{contnr}
\end{figure*}

\begin{figure*}
	\centering
	\includegraphics[scale=0.42]{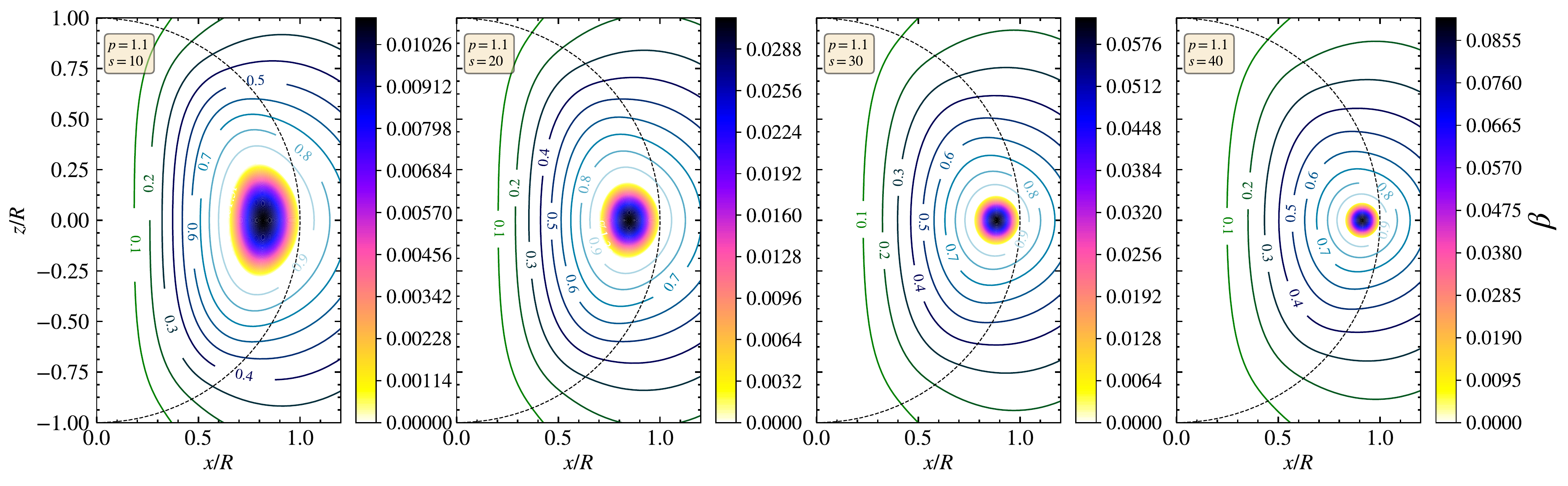}
	\includegraphics[scale=0.42]{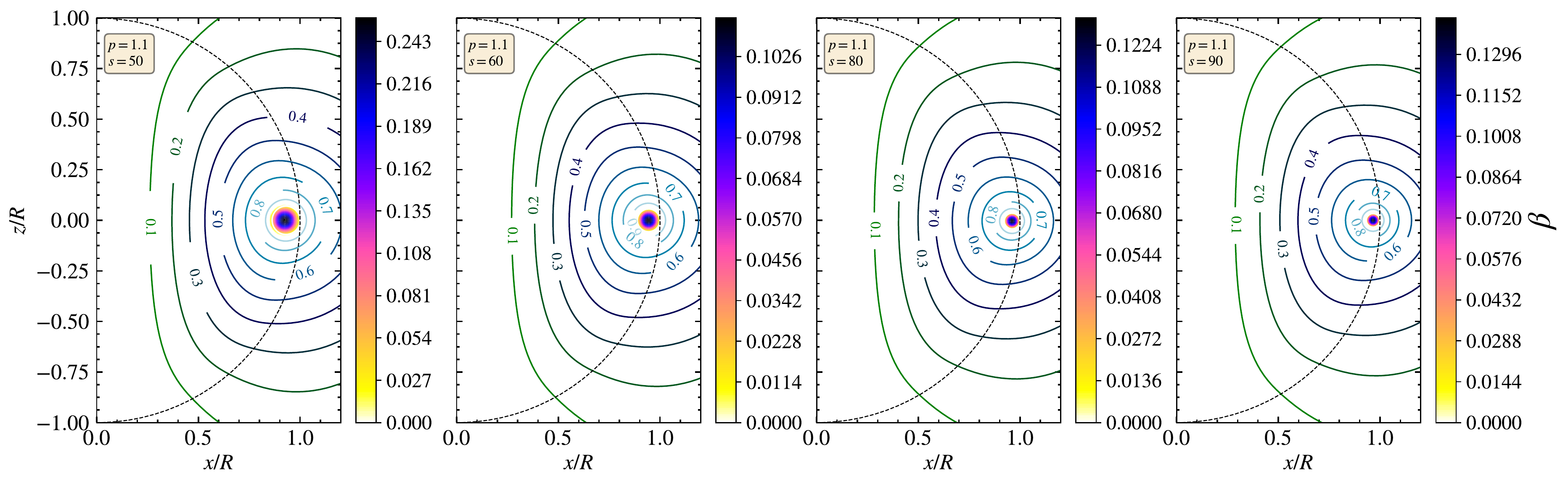}
	\caption[short]{Contours of poloidal field lines with different $s$ values for the $n=1$ polytropic density profile $\rho=\rho_c\frac{sin(\pi r/R)}{\pi r/R}$. The colorbar, again, shows the strength of $\beta$ and the red dotted line represent $r=R$. }
	\label{contrhon1}
\end{figure*}

{The age during which a realistic NS attends Hall/MHD equilibrium depends on its magnetic field strength ($B$). Typically, in a NS with an age of $10^3-10^5$ yrs having $B \geq 10^{14}$ G, the Hall term dominates over the Ohmic dissipation term. However, in a NS with and Ohmic timescale of a billion years, the Hall term can still dominate for weaker field strengths, for example in pulsars with $B\sim10^{12}$ G, as the internal temperature is lower and of the order $10^7-10^8$ K. Nore that although the Hall term is independent of the internal temperature ($T_{\rm in}$), the magnetic and thermal evolution of a NS are coupled \citep{Vigano2013}, as the magnetic diffusivity (which we neglect) is strongly dependent on the internal temperature. In practice for the realistic systems we consider, we require $T_{\rm in}\lesssim 10^{8}$ K, which also ensure the core temperature is well below the critical temperature for proton superconductivity.}\\

In this section, we discuss axisymmetric solutions for the three different models: a) Normal matter in the crust and the core - this includes both the case of standard MHD equilibria, and Hall equilibria, b) Hall in the crust and MHD in the core, and the more realistic case c) Hall equilibrium in the crust and a superconducting core. 

\subsection{Normal matter in crust and core}
\label{41}

In this section, we consider a star composed of normally conducting matter in both the crust and the core. We show results for the Hall equilibrium, however these results can be easily extrapolated to MHD equilibrium, by replacing $n_e$ with $\rho = \rho_c(1-r^2)$,  replacing $\chi_{\rm MHD} = Y_e \, \chi_{\rm Hall}$ and changing the constants given as $\lambda_{\rm MHD} \rho_cR^2 = B_0^{\prime}$.

In the following we choose $\chi_{\rm Hall}(\alpha) = \lambda_{\rm Hall} \alpha$ and set $\lambda_{\rm Hall}=10^{-35}$ G cm. This constant $\lambda_{\rm Hall}$ sets the strength of the magnetic field ($B_0$). As mentioned before, we present results for the Hall equilibria models with two different electron density profiles, one constant in space $n_1 = n_e$, and the other radially decreasing profile $n_2(r) = n_e(1-r^2)$ inside the NS. {We also show results for the EOS with matter density following the $n=1$ polytrope}. We solve equation \ref{eq1} with the $\alpha$ and $\beta$ normalized by $B_0/R^2$ and $B_0/R$ respectively. The normalization constant $n_e$ of the electron density profile, with typical values $10^{36}-10^{34}$ cm$^{-3}$ across the crust, appears as $\lambda_{\rm Hall}\,n_e\,R^2 = B_0.$ For a star with radius $R=10$ km, we get $ B_0\sim 10^{13}$ G {which corresponds to a surface field strength of $\sim10^{11} \rm G$. The results we obtain are scalable to any field strength $B_0$ which doesn't influence our magnetic field topology but results in changing the quadrupolar deformation which we calculate later in section \ref{magdef}.} The normalized GS equation for the Hall equilibria is given by
\begin{eqnarray}
\Delta^{\star} \alpha = -\bigg(\frac{\lambda_{\rm Hall}\,n_e \,R^2}{B_0}\bigg)r^2(1-\mu^2)n(r) - \beta\beta^{\prime}  \hspace{1cm} 
\end{eqnarray}
We consider now results for the whole star, i.e. $r_{\rm min}=0$, since it provides simpler analytical expressions.To compare with previous studies we choose $p=1.1$. A lower value of $p$, in principle, develops stronger toroidal field. However $p$ cannot be less than 0.5 as it makes the term $ \beta^{\prime}\beta$ infinite in certain regions inside the star \citep{2013MNRAS.434.2480G}. For $n_1=1$, a purely poloidal field ($s=0$) has an analytical solution given by \cite{2013MNRAS.434.2480G}
\begin{equation}
\alpha(r,\mu) = 
\begin{cases}
\frac{(1-\mu^2)}{30}(5r^2-3r^4)       & \quad \text{if } r  < 1\\
\frac{(1-\mu^2)}{15}\frac{1}{r}  &\quad \text{if } r\geq1
\end{cases}
\end{equation}

\begin{table}[h]
	\begin{center}
		
		\begin{tabular}{c|c|c|c} 
			\textbf{s} &\textbf{this work} & \textbf{Armaza} & \textbf{Gourgouliatos} \\
			\hline
			0 & 0 & 0 & 0\\
			5 & 0.12  & 0.14 & 0.15\\
			10 & 0.65 & 0.57 & 0.60\\
			20 & 2.2 & 2.2 & 2.3\\
			25 & 3.1 & 3.1 & 3.2\\
			30 & 3.3 & 3.7 & 3.9\\

		\end{tabular}
		\caption{The percentage of $\mathcal{E}_{tor}/\mathcal{E}_{mag}$ for different parameter values of $s$ for $n_2(r)=(1-r^2)$. A comparison is also shown with \cite{2015ApJ...802..121A} and \cite{2013MNRAS.434.2480G}.}
		\label{table1}
	\end{center}
\end{table}

\begin{figure}[!h]
	\centering
	\includegraphics[scale=0.45]{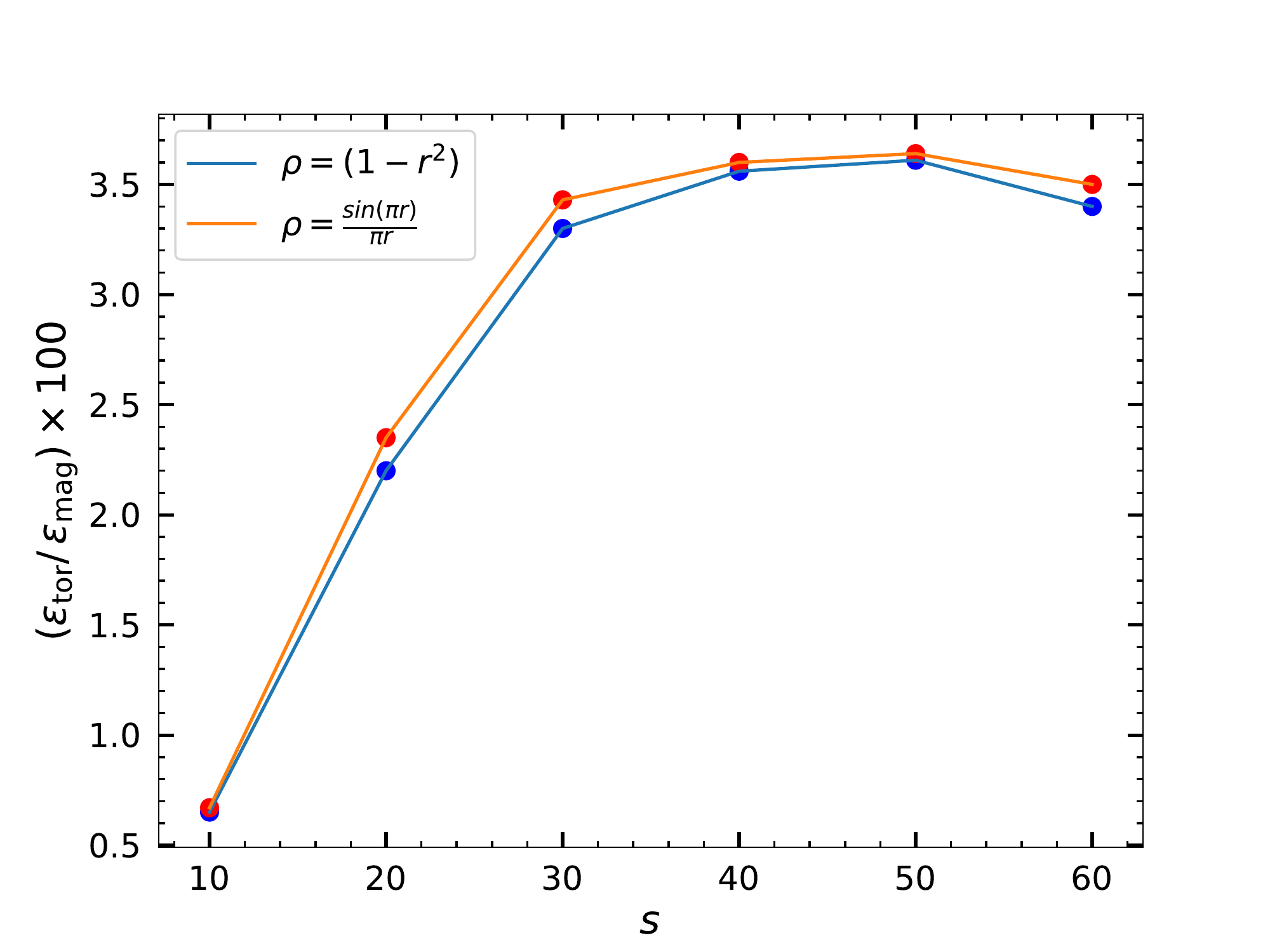}
	\caption{Percentage fraction of the toroidal magnetic energy ($\mathcal{E}_{tor}$) to the total magnetic energy $\mathcal{E}_{mag}$) for two different density profiles given in figure labels for the setup given in subsection \ref{41}.}
	\label{n1poly}
\end{figure}

\begin{figure*}
	\centering
	\includegraphics[scale=0.5]{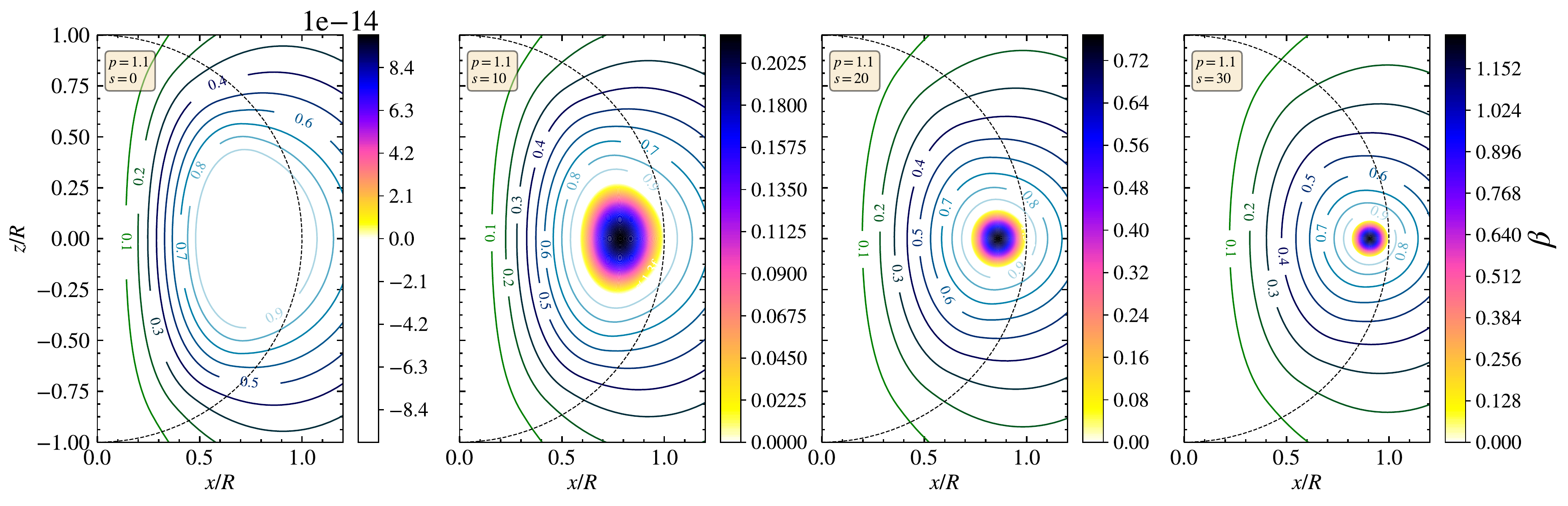}
	\caption{Magnetic field lines and the strength of $\beta$ for the Hall equilibrium in the crust and MHD equilibrium in core.}
	\label{gshm}
\end{figure*}

This is represented by the black line while the numerical calculations are shown by the green triangles in figure \ref{analy_sol}. In figure \ref{variation_alpha} , we compare the radial and angular variation of the poloidal field for the two electron density profiles. We show results for different values of $s \in [0,10,20,30,40,50,60,80,90]$. Firstly, varying the electron density yields a weaker poloidal field. Secondly, a higher value of $s$ makes the poloidal field stronger with its maximum value lying at the equator as seen in sub-figures \ref{sol1b}, \ref{sol2b} and \ref{sol3b}. This is in good agreement with the results obtained by \citet{2013MNRAS.434.2480G}. However, on increasing $s>80$, we do not see a further rise in the peak of poloidal field strength. Moreover, the results in figure \ref{sol2b} show qualitative convergence as we increase the value of $s$, from which we conclude that models with $s\sim50$ may be used as a reasonable approximation for the field structure. The geometry of the field lines for these different cases are shown in figures \ref{contn1},\ref{contnr} and \ref{contrhon1}, for $s \in [0-90]$, with the colorscale representing the strength of $\beta$ which is directly proportional to the toroidal field strength. The toroidal component is concentrated close to the stellar {equator} and lies along the neutral line where the innermost closed poloidal field line is located within the star. This is however not surprising since we chose to have no currents outside the star. These figures also show that increasing $s$ makes the region containing the toroidal field smaller as predicted by previous studies \citep{2009MNRAS.395.2162L,2013MNRAS.434.2480G,2015ApJ...802..121A}. The toroidal region is also larger for the radially varying density profile $n_2$ when compared with the constant profile $n_1$. We compare our results directly with \cite{2013MNRAS.434.2480G, 2015ApJ...802..121A}  in table \ref{table1} where we show the percentage of the toroidal magnetic energy ($\mathcal{E}_{tor}$) to the total magnetic energy ($\mathcal{E}_{mag}$) with a fixed background density. The energies are comparable which shows that our results are consistent.  

{On comparing figures \ref{sol2a} and \ref{sol3a}, we see that the maximum value of $\alpha$ across the equatorial region is lower for the density profile $\rho_2(r) \propto \frac{sin(\pi r)}{\pi r}$ when compared to  $n_2(r) \propto (1-r^2)$. This corresponds to a weaker poloidal and toroidal component however on comparing the fraction of toroidal energy with different values of $s$ as seen in figure \ref{n1poly}, we see that the energies are comparable which assures our assumption that both these density profiles resemble each other}.

We remark again that as we have assumed that density to be a function of radius only, the pressure and gravity forces are also radial and hence cannot balance the angular component of the magnetic force. In principle these forces will deform the star and lead to an ellipticity, and one should also solve the evolution equation for the density \citep{2009MNRAS.395.2162L}. However, for the magnetic fields in regular pulsars,  magnetic equilibrium can be treated as a perturbation on the background \citep{Akg2013}. Deformations of the density profile are of the  higher order in $B^2$, and any back reaction on the field is even smaller, O($B^4$), and hence will only play a role for very strong magnetic fields in magnetars. 

We have explored a different twisted-torus geometry with a continuous toroidal field $\beta(\alpha) = \gamma \alpha (\alpha/\bar{\alpha}-1)\Theta(\alpha/\bar{\alpha}-1)$ 
where $\bar{\alpha}$ is the value at the last closed-field line and $\gamma$ is a constant. This entire framework was carried out in general relativity by \citet{Ciolfi2013}(C\&R) which we try to reproduce in the Newtonian limit. Previously, we have seen that with increasing $s$, the toroidal field becomes stronger and the closed-field line region shrinks. To produce a larger closed-field line, C\&R considered a functional dependence of $\chi(\alpha) = c_0[(1-\lvert\alpha/\bar{\alpha}\rvert)^4\Theta(1-\lvert \alpha/\bar{\alpha}\rvert)-\bar{k}]$, with $c_0$ and $\bar{k}$ are constants. Furthermore the transformation $\chi(\alpha) = \chi(\alpha)+\bar{\chi}(\alpha)$ was applied, with $\bar{\chi}=X(\alpha)\beta^{\prime}\beta$, thus minimizing the effect of toroidal fields on the poloidal field lines. With $\gamma=1$, $c_0 = 1$, $\bar{k}=0.03$, and $X(\alpha)=1$ we solved the GS equation and get $\mathcal{E}_{tor}/\mathcal{E}_{mag} \sim 0.05$ instead of the very strong toroidal fields $\mathcal{E}_{tor}/\mathcal{E}_{mag} \sim 0.6$ obtained by C\&R \citep{Ciolfi2013}.{We do not solve our equations in GR and cannot conclusively say what could give rise to this discrepancy.

\subsection{Hall equilibria in crust and MHD in core}
\label{42}

As a first step towards more realistic models, we start by considering the case where we have a Hall equilibrium in the crust, and an MHD equilibrium in the core of the star. We follow \cite{Fujisawa2014} who showed that the strength and structure of magnetic field in the core affects that in the crust, and the current-sheet at the crust-core interface affects the internal and external field. Similarly, we look into a situation where we impose Hall equilibrium in the crust and MHD equilibrium in the core. {Outside the star, we assume vacuum condition, in which case, we solve $\Delta^{\star} \alpha=0$ with the zero boundary conditions for $\alpha$ at a far away radial point. One can also impose a dipolar field however the results do not change significantly as seen by \cite{2013MNRAS.434.2480G}.}
  
\begin{figure}[!h]
	\centering
	\includegraphics[scale=0.46]{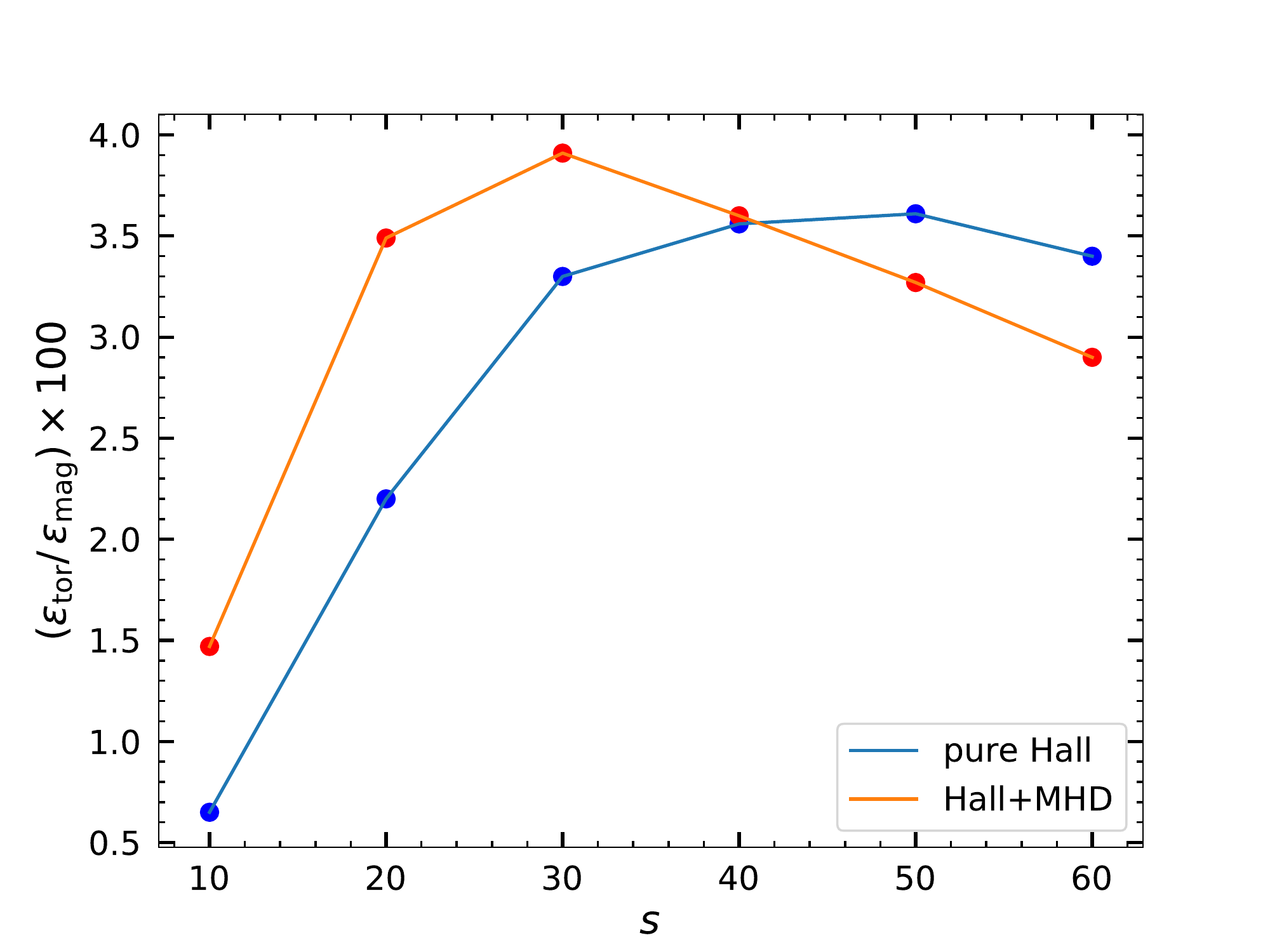}
	\caption{Percentage fraction of the toroidal magnetic energy ($\mathcal{E}_{tor}$) to the total magnetic energy $\mathcal{E}_{mag}$) for the pure Hall (setup in \ref{41}) and mixed Hall+MHD (setup in \ref{42}) with varying $s$.}
	\label{ratio}
\end{figure}

In order to study this case we now have to, unlike in previous examples, explicitly differentiate between Hall and MHD equilibria. The equations we solve are given by
\begin{align}
&\Delta^{\star} \alpha = -r^2(1-\mu^2)n(r)\chi^\prime_{\rm Hall} - \beta\beta^{\prime}  \hspace{1cm} \rm in \, \, crust\\
&\Delta^{\star} \alpha = -r^2(1-\mu^2)\rho(r)\chi^\prime_{\rm MHD} - \beta\beta^{\prime} \hspace{1cm} \rm in \, \, core
\label{hallmhd}
\end{align}
At the crust-core interface, the continuity of $\alpha$ is automatically imposed. We also want the magnetic field in the core and the Lorentz force in the crust to balance, which gives 
\begin{eqnarray}
\bigg[\rho_{\rm core} \,\chi^{\prime}_{\rm MHD}\bigg]^{cc} = \bigg[n_{\rm crust} \,\chi^{\prime}_{\rm Hall}\bigg]^{cc}
\label{ccbalance}
\end{eqnarray}
We set the electron density in the crust to be a constant $n_{\rm crust}=n_e$ while the density in the core is assumed to follow $\rho_{\rm core} = \rho_c(1-r^2)$. With the crust-core boundary at $r=0.9R$, using equation \ref{ccbalance} we get the following relation
\begin{eqnarray}
\chi^{\prime}_{\rm MHD} = 5.1 \times 10^{21} \chi^{\prime}_{\rm Hall} \sim 5.1 \times 10^{-14} {\rm \,G \,cm^{-1}}
\end{eqnarray}
The magnetic field lines remains unchanged however the strength of the parameter $\beta$ is significantly higher as seen in figure \ref{gshm}. 

We plot the percentage fraction of toroidal energy for the pure Hall+MHD in fig \ref{ratio} and compare this with the pure Hall equilibrium NS. The difference between this setup compared to purely MHD or Hall equilibria is that the toroidal energy density is stronger up to $s\sim 40$, but starts decreasing for higher values. Qualitatively, however, the results are similar and the toroidal energy saturates at a few percent of the total energy.

\begin{figure*}[h]
	\centering
	\begin{subfigure}{.5\textwidth}
		\centering
		\includegraphics[scale=0.45]{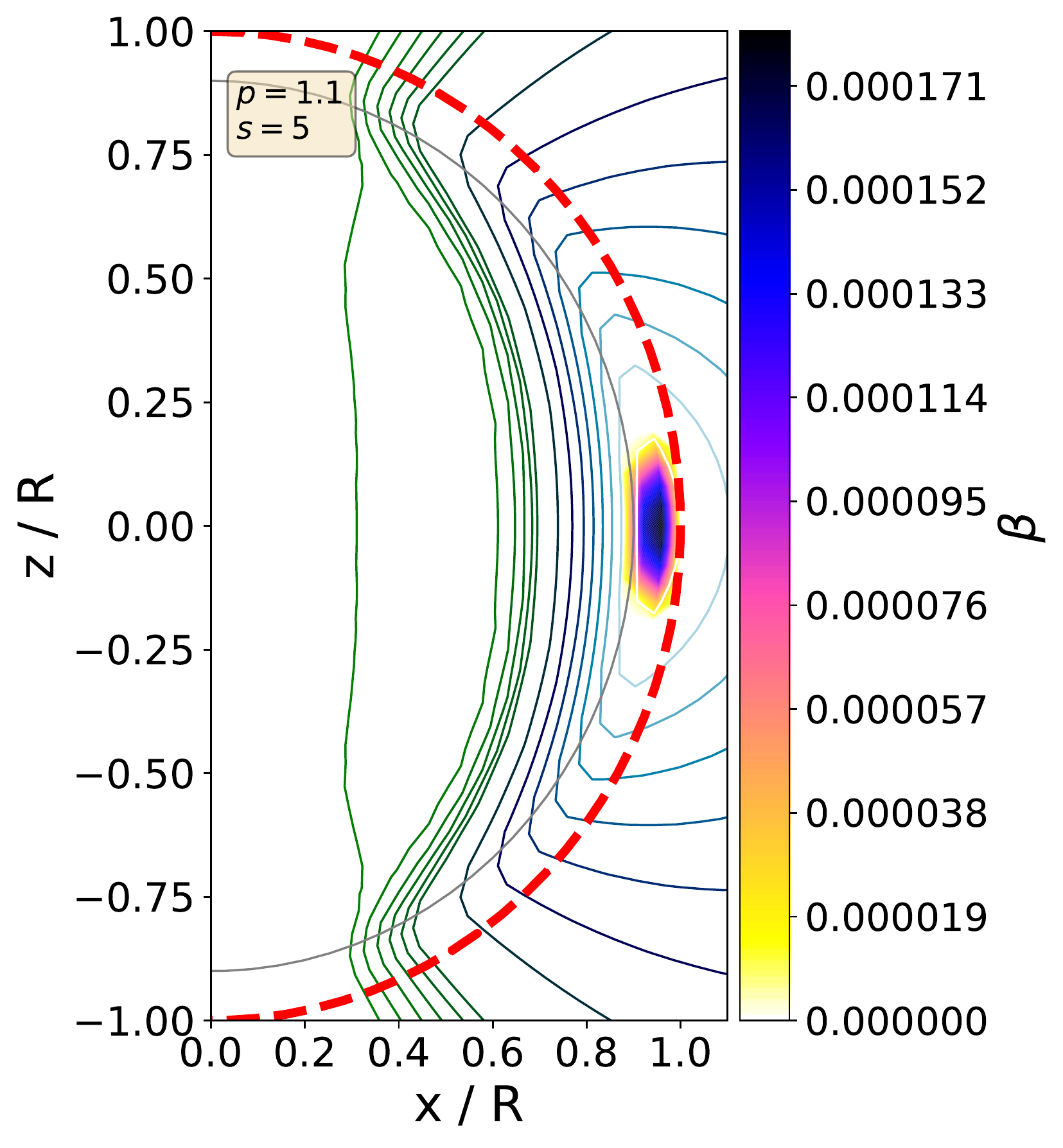}
		\caption{$s=5$}
		\label{sc5}
	\end{subfigure}%
	\begin{subfigure}{.5\textwidth}
		\centering
		\includegraphics[scale=0.45]{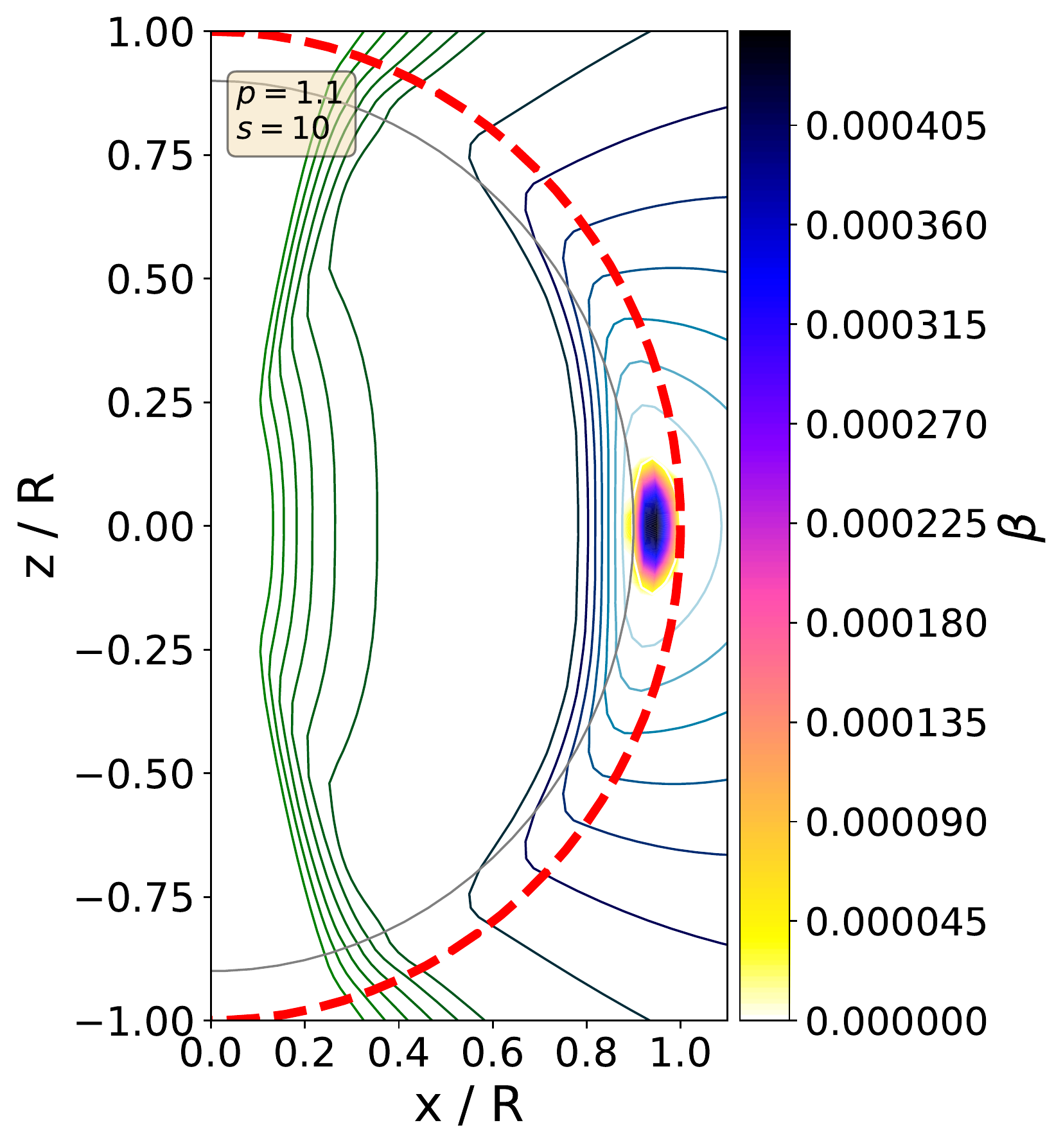}
		\caption{$s=10$}
		\label{sc10}
	\end{subfigure}    
	\caption{Contours of poloidal field lines for two different values of $s$ in the superconducting core-Hall equilibrium crust. The location of the crust-core interface is represented by the solid gray line at $r=0.9R$, while the red-dotted line shows the stellar surface. The colorscale represents the strength of the toroidal field $\beta$. }
\end{figure*}

\subsection{Hall equilibrium in crust and Superconducting core}

As the NS cools down, neutrons and protons form pairs by reducing their energy owing to the long range attractive part of the residual strong interactions. The thermal energy in this case is much smaller than the pairing energy and the system is gaped, which leads to reactions and viscosity being greatly suppressed. As previously mentioned, the transition temperature below which the system behaves as superfluid/superconductor is typically $T_c \sim 10^{9}-10^{10}$ K, and the star thus cools below this rapidly after birth. In the interior of a mature neutron star the geometry of the magnetic field depends on the type of superconductivity, which in turn depends on the size of Cooper pairs and the penetration depth of the magnetic field. This is measured with the Ginzburg-Landua parameter $\kappa_{GL}$ which for NS cores is greater than $1/\sqrt{2}$ making it a type-II superconductor (for an in depth discussion, see the review article by \citet{HaskellSedrakian}).\\
	
In order to obtain a more realistic model of a pulsar, we thus consider a core of type II superconducting protons, and study its effect on the magnetic field equilibria, following the setup of \cite{Lander2013PRL,Lander2014}. Therefore, our star is made up of normal matter in Hall equilibrium in the crust and superconducting matter in the core. The crust-core interface lies at $r=0.9$ as in previous cases. {Our models, in this case, are valid for pulsars having a surface field strength of $(1\--10)\times 10^{11}$ G with typical ages in the range of $10^{4}-10^{5}$ yrs and internal temperatures of $10^7-10^9$ K.} The Lorentz force for this type-II superconducting protons in the core is given by \citep{Easson,Mendell1991,Akg2008,Kostas2011}

\begin{equation}
\vec{F}_{\rm mag}  = -\frac{1}{4\pi}\bigg[\vec{B}\times(\vec{\nabla}\times \vec{H}_{cl}) + \rho_p \vec{\nabla}\bigg(B\frac{\partial H_{c1}}{\partial \rho_p}\bigg)\bigg]
\label{LFsc}
\end{equation}
where $\vec{H}_{cl}(\rho_p,\rho_n) = H_{c1} {\hat{B}}$ is the first critical field with ${\hat{B}}$ is the unit vector tangent to the magnetic field. The norm of this first critical field is given as
\begin{eqnarray}
H_{c1}(\rho_n,\rho_p) = h_c\frac{\rho_p}{\varepsilon_{\star}}
\end{eqnarray}
{where} $h_c$ is an arbitrary constant \citep{Kostas2011}. {We assume that the density of protons $\rho_p$ in the core follows the same profile $\approx (1-r^2)$ as that by electrons in the crust.} The entrainment parameter is given by $\varepsilon_{\star} = \frac{1-\varepsilon_{p}-\varepsilon_{n}}{1-\varepsilon_{n}}$, where $\varepsilon_{p} = 1-\frac{m_p^{\star}}{m_p}$. Here $m_p^{\star}$ is the effective mass of the protons acquired as a result of entrainment. Similarly, we can define $\varepsilon_{n}$. We refer the reader to \cite{Palapanidis2015} where the effect of entrainment is discussed extensively. In the following, we simply set $\varepsilon_{\star}=1$, which implies that force on neutrons due to coupling is zero which allows us to represent $H_{c1} = H_{c1}(\rho_p)$. The equivalent Grad-Shafranov equation for type-II superconducting core is thus given by
\begin{eqnarray}
\Delta^{\star}\alpha = \frac{\vec{\nabla}\Pi\cdot \vec{\nabla}\alpha}{\Pi} - r^2(1-u^2)\rho_p\Pi\frac{dy}{d\alpha} - \Pi^2 f_{sc}\frac{df_{sc}}{d\alpha}
\label{gs_sc}
\end{eqnarray}
where we represent superconducting matter with the subscript $sc$ and  the functions $f_{sc}$ and $y(\alpha)$ are defined as
\begin{align}
&y(\alpha) = 4\pi \chi_{sc} + B\frac{h_c}{\varepsilon_{\star}}\\
&f_{sc}(\alpha) = \frac{\beta}{B}H_{c1}
\end{align}
where $B = \sqrt{\vec{B}\cdot\vec{B})}$ is the magnitude of the magnetic field and $\Pi = \frac{B}{H_{c1}}$. Equation \ref{gs_sc} is valid in the NS core ($r<0.9R$). For the crust, we consider normal matter in Hall equilibrium while the exterior remains the same as considered before. As previously remarked, we consider mostly the more realistic case of Hall equilibria, but our equations in general can be applied also to MHD equilibrium, in which case the boundary conditions are modified. In particular equation \ref{eqfory} becomes:

\begin{eqnarray}
y(\alpha) = \frac{h_{c}}{\epsilon_{\star}}B_{\rm cc}(\alpha)  + 4\pi\bigg[\frac{\rho_p^{\rm crust}}{\rho_p^{\rm core}}\bigg]_{cc}\chi_{\rm MHD}(\alpha)
\label{yeq_bc}
\end{eqnarray}

\subsubsection{Boundary conditions}

{We treat the boundary conditions as outlined in \citep{Lander2014}. At the surface of the NS, the density of protons vanishes. The magnetic field in the core and the Lorentz force in the crust must balance, which gives} 
\begin{eqnarray}
\bigg[\rho_{p}^{\ core} \,\chi^{\prime}_{\rm sc}\bigg]^{cc} = \bigg[n_e^{\rm crust} \,\chi^{\prime}_{\rm Hall}\bigg]^{cc}
\label{cc_boun_sc}
\end{eqnarray}
Since apriori, we do not know $B$ explicitly as a function of $\alpha$, we use a polynomial approximation for $B$ at the crust-core interface as given in \citep{Lander2014}

\begin{eqnarray}
B_{\rm cc}(\alpha) = c_0 + c_1 \alpha + c_2\alpha(\alpha-\alpha_{\rm cc}^{\rm eq})
\end{eqnarray}
where $c_0$ are constants and $\alpha_{\rm cc}^{\rm eq}$ is the equatorial value of $\alpha$ on the crust-core boundary. We choose the constant such that

\begin{align}
c_0 &= B_{\rm cc}^{\rm pole} \\
c_1 &= \frac{B_{\rm cc}^{\rm eq}-c_0}{\alpha_{\rm cc}^{\rm eq}}\\
c_2 &= \frac{B_{\rm cc}^{\rm mid}-c_0 - c_1\alpha_{\rm cc}^{\rm mid}}{\alpha_{\rm cc}^{\rm mid}(\alpha_{\rm cc}^{\rm mid} - \alpha_{\rm cc}^{\rm eq})}
\end{align}
where $\alpha_{\rm cc}^{\rm mid}$ is the value of $\alpha$ at $\theta=\pi/4$ in the crust-core boundary. This gives

\begin{equation}
y(\alpha) = \frac{h_{c}}{\epsilon_{\star}}B_{\rm cc}(\alpha)  + 4\pi\bigg[\frac{n_e^{\rm crust}}{\rho_p^{\rm core}}\bigg]_{cc}\chi_{\rm Hall}(\alpha)
\label{eqfory}
\end{equation} 
The next boundary condition that we must satisfy is the continuity of $B_{\phi}$ which is given by
\begin{eqnarray}
f_{\rm sc}(\alpha) = [H_{\rm c1}]_{\rm cc}\frac{\beta(\alpha)}{B_{\rm cc}(\alpha)}
\end{eqnarray}

\subsubsection{Field lines}

The poloidal field contours are shown in figures \ref{sc5} and \ref{sc10} with the colorscale again representing the strength of $\beta$. {This corresponds to a maximum toroidal field of magnitude $10^{10}$ G. In the core, we see that the field lines are convex} for the superconducting matter as opposed to the normal matter. The toroidal field is also restricted to the crust and cannot penetrate deep within the star. This can be understood by comparing the ratio of averaged magnetic field strength to the magnitude of $H_{c1}$ at the crust-core interface i.e. $\langle B^{cc} \rangle/H_{c1}^{cc} < 1$, which is typically the case for pulsars. $\langle B^{cc} \rangle/H_{c1}^{cc} \geq 1$ makes the field lines kink inwards and close inside the core \citep{Lander2014}. This effect is independent of the choice of our function $\chi_{\rm MHD}(\alpha)$. {In this study, we typically have $H_{c1}$ 10-50 times stronger than the magnitude of $B$ at the crust-core interface, which increases the magnetic tension towards the $z$ axis. The toroidal flux is fully expelled to the crust, as also seen by \citet{Lander2014}, while magnetothermal evolutions by \citet{Elfritz2016} found on the contrary a toroidal field in the core. We note however that this result depends strongly on the intial conditions for the evolution in the core, and further analysis of their compatibility with the equilibria found here would be needed to obtain a full physical understanding of this discrepancy .}

\subsection{Magnetic deformation}
\label{magdef}
Finally, let us discuss the magnetic deformation of the star, which plays an important role in estimating the strength of gravitational radiation from NSs \citep{Ushomirsky2000, Haskell2008,Mastrano2013, Lasky2015, Gao2017,Magda2019,Chandra2020}. In our setup, as already discussed, this can be treated as a higher order effect in an expansion in $O(B^2)$. The strategy is thus to compute the magnetic field on a spherical background, as we have done in the previous section, and then evaluate the deformations of the density profile at $O(B^4)$. Following \citet{Haskell2008}, the theta component of the Lorentz force ($L_{\theta}$) term is given by:
\begin{eqnarray}
(\delta p + \rho \delta \Phi)\frac{d Y^0_2}{d \theta} = r\frac{[(\vec{\nabla} \times \vec{B} )\times \vec{B}]_{\theta}}{4\pi} = \frac{r L_{\theta}}{4 \pi}
\end{eqnarray}
Where $Y_2^0$ is the $m=0$ spherical harmonic. We further impose the Cowling approximation which gives $\delta \Phi = 0$ and on using the EOSs considered in this work, we calculate the quadrupole moment, following \cite{Ushomirsky2000}, as
\begin{eqnarray}
Q_{lm} = \int_{0}^{R} \delta \rho_{lm} (r) r^{l+2}dr
\end{eqnarray}
which on dividing by the z-th component of the moment of inertia (I$_{zz}$), gives us the ellipticity parameter $\in \,= \,Q_{20}/I_{zz}$. Note that our models are axisymmetric, and would not lead to gravitational wave emission, even in the presence of a significant deformation. However, if the magnetic axis is not aligned with the rotational axis, the deformation will not be axisymmetic and there will be components of the quadrupole also with $m\neq 0$, leading to emission at both the rotational frequency and twice the rotational frequency of the star \citep{BG96}. In order to obtain an estimate of the ellipticity, we thus make the standard approximation that $Q_{20}\approx Q_{01}$, neglecting geometric factor that depend on the inclination angle.

We can now compare the quadrupoles obtained for our setups with Hall equilibrium in the crust, but with MHD and superconducting cores. The results are very similar for the densities $\rho \sim (1-r^2)$ and $\rho \sim \sin(\pi/r)/\pi r$,  and for $s=10$, we obtain for the MHD core $\epsilon\approx 2\times 10^{-12}$, corresponding to an average poloidal field of $B_p = 6\times 10^{11}$ G (with a surface value of $B_s=2\times 10^{11}$ G) and toroidal field of $ B_t=5\times 10^{10}$ G, which is in line with theoretical expectations. For our setup with a {superconducting} core we obtain, {taking} $H_c \sim 10 B_0$, values of $\epsilon\approx 7\times 10^{-11}$, for $B_p = 4.3\times 10^{11}$ G, $ B_t=3\times 10^{10}$ G and a surface field of $B_s=3\times 10^{11}$ G.

This is significantly lower than the results obtained by \citet{Mastrano2016} who found $\in \sim 10^{-6}$ from spot-like magnetic field structures present in the crust due to Hall effect causing density perturbations for field strengths higher than $B \geq 10^{14}$ G. This difference is likely to be caused by the interplay between the overall stronger poloidal field in the core of the star and the (locally) strong toroidal field in the crust which compensate each other in our model, while \citet{Mastrano2016} consider fields only in the crust of the star, and non-barotropic equations of state. Nevertheless, a full magnetothermal evolution of the couple crust-core system would be needed to conclusively shed light on the issue.

\begin{figure*}[h!]
	\centering
	\begin{subfigure}{.5\textwidth}
		\centering
		\includegraphics[scale=0.46]{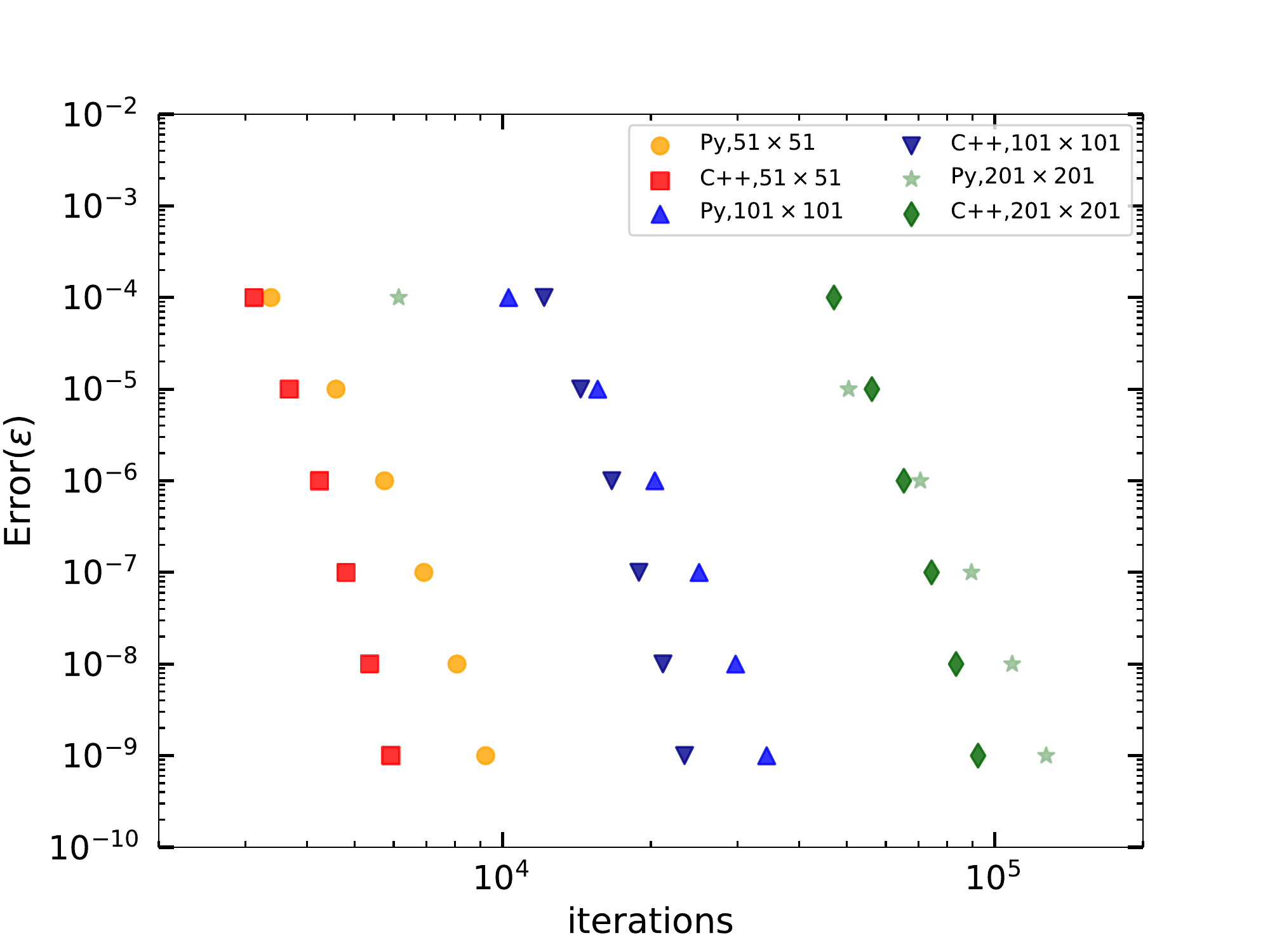}
		\caption{}
		\label{conv2}
	\end{subfigure}%
	\begin{subfigure}{.5\textwidth}
		\centering
		\includegraphics[scale=0.46]{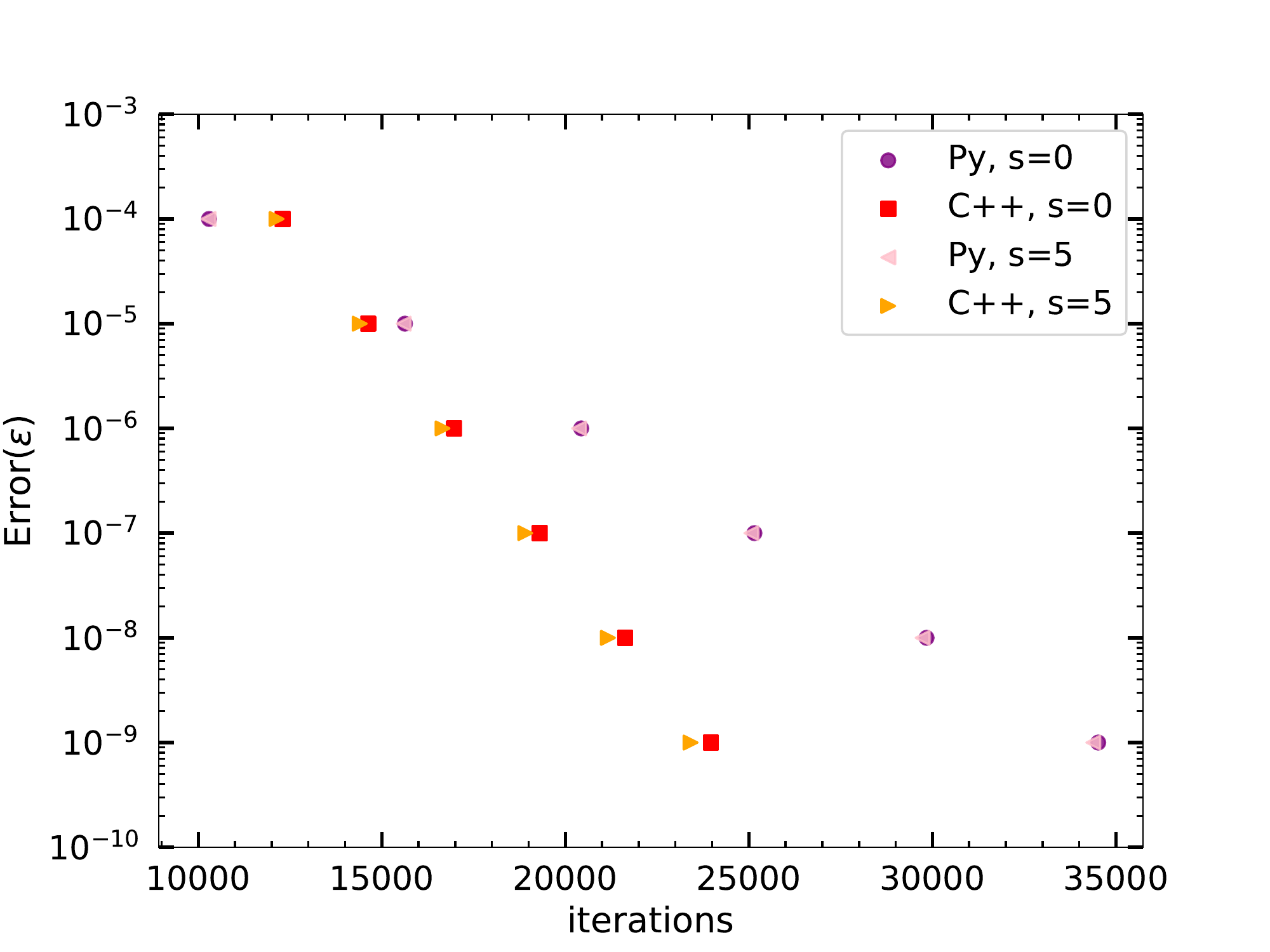}
		\caption{}
		\label{conv1}
	\end{subfigure}
	\caption{(a) Number of iterations as a function of accuracy for three different grid sizes for the two versions of our code. (b) The number of iterations taken by our solvers to reach a certain accuracy when generating different field geometries, purely poloidal ($s=0$) and the mixed poloidal-toroidal($s=5$). }
\end{figure*}

Note finally that in the MHD model the toroidal field regions of the star can present locally large deformations of density (up to $\delta\rho/\rho\approx 0.01$), which could be important in older, accreting systems, as if they occur in the crust they could lead to deformed capture layers in the presence of accretion \citep{Singh20}. However, in our more realistic models with a superconducting core, these deformations are much smaller, and never larger than $\delta\rho/\rho\approx 10^{-6}$.

\subsection{Code performance}

We compare our two codes in Python and C++. In figure \ref{conv2} , we show the number of iterations taken for each code to reach an error ($\varepsilon$) for three different grid sizes for $s=5$. In figure \ref{conv1}, we plot error as a function of number of iterations for two different cases, purely poloidal ($s=0$) and a mixed poloidal-toroidal ($s=5$). The grid size chosen was $101\times101$. Overall we infer that the performance of C++ is better as it takes less number of iterations (and hence time) to reach our final tolerance. However, for the purely poloidal case, we see that our Python code, which uses \texttt{numpy} vectorization, is faster when compared to C++. Further we calculated the order of convergence $oc = \ln\bigg(\frac{f_3-f_2}{f_2-f_1}\bigg)/\ln(r)$, where $f3, f2$ and $f1$ are values at a fixed point in the grid with resolutions $128\times128$, $64\times64$ and $32\times32$ respectively. Here $r$ is the refinement ratio chosen to be 2, and we find $oc\sim2$, showing second order convergence ($\mathcal{O}(h^2)$).

\section{Conclusion and Discussions}

In this paper we have developed a numerical scheme to rapidly solve the GS equation to obtain axisymmetric magnetic field equilibrium models for mature neutron stars.
As a benchmark for our code we first consider first the case of pure MHD equilibria and pure Hall equilibria, then move on to models in which we solve for Hall equilibrium in the crust and MHD equilibtium in the core. Finally, we produce for the first time a model where we consider a type II superconducting core and Hall equilibrium in the crust, thus producing a more realistic model for a mature pulsar.\\

We compared our numerical computations with analytical solutions provided by \citet{2013MNRAS.434.2480G} and with the results obtained by \citet{2015ApJ...802..121A} which shows excellent agreement. Since our source terms have a high-degree of nonlinearity, we have implemented a new technique which allows us to linearize our source. This, along with the under relaxation to update $\alpha$ had significantly improved our solver's performance. We were able to extend calculations for $s>65$, i.e. regions where previous studies \citep{2015ApJ...802..121A,2013MNRAS.434.2480G} had failed, for the normally conducting fluid. We reach convergence when $s\sim 50$ and the results do not change significantly beyond this value of $s$. However, our code fails when we increase $s$ beyond 90 for $p=1.1$ because the toroidal field region became too small to be resolved, causing numerical instabilities. We can do better if $p$ is increased to $2$, but this does not produce any difference in the toroidal component. In our calculations we implement both simplified equations of state, such as a constant density profile, in order to compare to previous results, but also more realistic profiles. In particular we implement a parabolic equation of state of the form $\rho\propto (1-r^2/R^2)$ and compare the results to those obtained with an $n=1$ polytrope, and find them to be in good agreement.\\

In this work, we have assumed the star to be always spherical. There are a few limitations of this as our results do not account for the back-reaction of the magnetic force on the fluid, and so we cannot self-consistently calculate ellipticities for strong magnetic fields, such as those of magnetars, where our perturbative approach is no longer valid. To consistently account for the effect of non-sphericity in the presence of strong fields or rotation, it would be necessary to modify our method to calculate simultaneously the density and field structure \citep{2009MNRAS.395.2162L}. Nevertheless our method can be confidently applied to the standard pulsar population, in which the magnetic fields are weak enough to enable a perturbative treatment, and rotationally induced deformations can safely be ignored. \\

We calculate magnetic equilibria solutions for the superconducting core NSs which differed from the normally conducting matter by field strengths and geometry of the poloidal fieldlines. The most notable change is that the toroidal field is expelled from the core and restricted to the crust in the case of where $\langle B^{cc} \rangle/H_{cl}^{cc}< 1$, which is applicable for the standard pulsar population. Increasing this ratio beyond 1 (as would be the case for magnetars) caused numerical difficulties and we leave it as a scope for improvement in the future.  Furthermore for $B>>H_c$ even nonlinear effects such as those considered bu \cite{2009MNRAS.395.2162L} can be important, and further development of our scheme would thus be required to consistently describe magnetars. \\

{To be precise, our Hall crust-superconducting core models are valid for middle-aged pulsars with ages of around $10^{5}$ years and core temperature of $T\lesssim 10^{9}$ K. We consider surface values of the magnetic field os order $B_s\leq10^{12}$ G. This guarantees that we are exploring a population which have a timescale for Hall evolution which is much shorter than the Ohmic dissipation timescale and we can ignore the latter contributions when computing our models.}\\

Our results for superconducting cores and Hall equilibrium in the crust are, therefore, a realistic model for mature pulsars, and are particularly interesting for glitching pulsars.
It has, in fact, been suggested that a strong toroidal field region in the core could lead to vortex/flux tube pinning, thus providing a large reservoir of angular momentum to power large glitches such as those observed in the Vela pulsar \citep{Erbil17, Erbil20}, and possibly resolving the tension between the observed activity of the Vela (i.e. the amount of spin-down reversed by glitches during the observing period) and the angular momentum that theoretical models predict to be stored in the crust \citep{crust12, chamel13}. 
Our model, however, shows that no toroidal field area is present in the core to allow for such pinning, as it is expelled to the ``normal'' matter crust. The models we produce can, however, be used as a background for more realistic vortex pinning calculations \citep{Sourie1, Sourie2}, in order to fully investigate the effect of pinning in the core on pulsar glitch phenomenology. {Note, however, that for very strong surface magnetic fields (e.g. $B_s>10^{15}$ G) like those seen in magnetars, the ratio $\langle B^{cc} \rangle/H_{c1}^{cc}$ is greater than unity and the toroidal flux is non-zero inside the core of the star \citep{Lander2014}}\\

All our computations produced a toroidal field which is less than $5 \%$ of the total magnetic energy, and both the structure and strength of the field appear to rapidly converge to a qualitatively stable regime as we increase the degree of nonlinearity by increasing the parameter $s$. This is in line also with the results obtained from numerical MHD evolution by \citet{2020MNRAS.495.1360S}. We have also shown that we can have a stronger toroidal field for a model with Hall equilibrium in crust and MHD equilibrium in the core of the star, and also tried to implement the formalism presented in \cite{Ciolfi2013} to generate extremely strong toroidal fields. We do not obtain results with toroidal energies significantly larger than $\simeq 5\%$ of the total magnetic energy of the star, although we do not work in GR and do not consider strong magnetic fields in our setup. This has strong implications as the gravitational-wave emission, as the size of a `mountain', i.e. of the quadrupolar deformation that couples to the gravitational field, strongly depends on the internal magnetic energy of a NS \citep{Haskell2008}.\\

We study such `mountains' on  the star by calculating the density and pressure perturbation induced at higher order by the field configurations we generate. This allows us to estimate the ellipticity, which in our case is $\in \approx 7\times 10^{-11}$ in the superconducting case, for a surface field value of $B_s\approx 3\times 10^{11}$ G, which is in line with theoretical expectations and confirms that if such deformations persist also in older pulsars, for which the crustal field may be buried leading to a lower inferred external dipole, this may explain the observed cutoff observed in the $P-\dot{P}$ diagram for millisecond pulsars by \citet{Woan18}.\\

We have written two versions of our code, one in C++ and the other in Python. To improve our python code's performance, we vectorized our arrays to perform operations instead of using loops. The number of grid points play a major role in resolving the toroidal component. We needed finer mesh to obtain the strong toroidal fields wherein our C++ code was efficient. Using our python code, we could solve the GS equation with parameters ($s=0, \, p=1.1$) for $101\times101$ grid to reach an accuracy of $\epsilon \sim 10^{-8}$ in less than 10 seconds. With the nonlinear source, the python solver took longer time and hence we used our C++ solver. \\

To summarize, we generated realistic magnetic equilibrium models in NSs with superconductiong cores, which could serve as initial conditions for long-term evolution of the magnetic field. This will bestow our understanding of the global magnetic field structure and its stability over the lifetime of a NS or any barotropic star.

\section{Acknowledgement}
AS and BH thanks Samuel Lander for carefully reading the manuscript and providing valuable suggestions. AS also thanks Sandip Mazumdar for pointing out the under-relaxation scheme mentioned in his book. 

\section{Financial Support}
AS was supported by the OPUS grant from the National Science Centre, Poland (NCN), number 2018/29/B/ST9/02013. BH was supported by the grant number 2015/18/E/ST9/00577. 

\section{Conflicts of Interest}
None

\bibliographystyle{pasa-mnras}
\bibliography{refs}

\end{document}